\documentclass[a4paper]{article}
\usepackage{amssymb}
\usepackage[dvips]{graphicx}
\usepackage{hyperref}

\newcommand{\R}{\mathbb{R}}
\newcommand{\Z}{\mathbb{Z}}
\newcommand{\N}{\mathbb{N}}
\newcommand{\Q}{\mathbb{Q}}
\newcommand{\K}{\mathrm{K}}
\newcommand{\cn}{\mathrm{cn}}

\newcommand{\qed}{\hspace{\stretch{1}}$\square$\\}

\newtheorem{theorem}{Theorem}
\newtheorem{lemma}{Lemma}
\newtheorem{corollary}{Corollary}
\newtheorem{proposition}{Proposition}
\newtheorem{remark}{Remark}

\begin{document}
\title{Chaotic quasi-collision trajectories in the 3-centre problem}
\author{L. Dimare\footnote{Dipartimento di Matematica, Universit\`a di
    Roma `La Sapienza', P.le Aldo Moro, 2, 00185-Roma, Italy; e-mail:
    {\tt dimare@mat.uniroma1.it} }} 
\date{}

\maketitle 
\begin{abstract}
We study a particular kind of chaotic dynamics for the planar
3-centre problem on small negative energy level sets.
%
%
We know that chaotic motions exist, if we make the assumption that one
of the centres is far away from the other two (see \cite{BN2}): this
result has been obtained by the use of the Poincar\'e-Melnikov theory.
Here we change the assumption on the third centre: we do not make any
hypothesis on its position, and we obtain a perturbation of the
2-centre problem by assuming its intensity to be very small. Then, for
a dense subset of possible positions of the perturbing centre in
$\R^2$, we prove the existence of uniformly hyperbolic invariant sets
of periodic and chaotic almost collision orbits by the use of a
general result of Bolotin and MacKay (see \cite{BM1}, \cite{BM2}).  To
apply it, we must preliminarily construct \emph{chains of collision
  arcs} in a proper way.  We succeed in doing that by the classical
regularisation of the 2-centre problem and the use of the periodic
orbits of the regularised problem passing through the third centre.

\vskip 0.3cm 
\noindent
{\bf keywords:} Centre problem, Regularisation, Collisions,
  Chaotic motion.
\end{abstract}

\section{Introduction}
We consider the motion of a particle in the plane, under the gravitational 
action of three  point masses at fixed positions (the planar restricted 
3-centre problem). We fix a Cartesian reference system $Oxy$ on the plane 
and choose suitable dimensionless coordinates such that the two centres 
with greater masses occupy the positions $C_1=(1,0),C_2=(-1,0)$. Following 
the common terminology, we refer to these as primaries. We suppose for 
simplicity that the primaries have equal intensities 
$a_1=a_2=a>0\,$ (symmetric problem),
and we always consider all the three centres having positive intensities. 

Let $C=(x_0,y_0)\in \R^2\setminus\{C_1,C_2\}$ be the position of the third 
centre and $\varepsilon >0$ be its intensity. We assume $\varepsilon$ to be 
very small and consider the limit $\varepsilon \rightarrow 0$: this means that 
$\varepsilon$ is a perturbation parameter.  In other words, we  make the 
hypothesis that the conditions are such that we can consider the problem as a 
one-parameter perturbation of an integrable one: the 2-centre problem. 

Let $M$ be the smooth Riemannian manifold
$M=\R^2\setminus\{C_1,C_2\}$, with the induced Euclidean metric on the
tangent bundle $TM$. The configuration space for the motion of a
particle in the potential field generated by the centres $C_1,C_2,C$,
is $M \setminus \{C\}$.  Let $P=(x,y)$ denote the position of the
particle on $M\setminus\{C\}$.  Then the system has smooth Lagrangian
function on $T(M\setminus\{C\})$ given by
\[
L_{\varepsilon}=L_0+\frac{\varepsilon}{\sqrt{(x-x_0)^2+(y-y_0)^2}}\,,
\]
with $L_0$ the Lagrangian of the 2-centre problem
\[
L_0=\frac{\dot{x}^2+\dot{y}^2}{2}
+\frac{a}{\sqrt{(x+1)^2+y^2}}+\frac{a}{\sqrt{(x-1)^2+y^2}}\ .
\]
Here we use the Newtonian notation for derivatives with respect to time: 
$\dot{x}=dx/dt, \dot{y}=dy/dt$. Note that $L_0$ is a smooth function on $M$, 
while $L_{\varepsilon}$ has a Newtonian singularity at the point $C$. 
To simplify notation, let us denote by $W(x,y)$ and $\varepsilon V(x,y)$ the 
potentials due respectively to the primaries and the third centre $C$, so that
\begin{eqnarray*}
  W(x,y)&=&-\frac{a}{\sqrt{(x+1)^2+y^2}}-
  \frac{a}{\sqrt{(x-1)^2+y^2}}\,,
  \\
  V(x,y)&=&-\frac{1}{\sqrt{(x-x_0)^2+(y-y_0)^2}}\ .
\end{eqnarray*}
The Lagrangian and the Hamiltonian of the problem have the form
\[
L_{\varepsilon}=L_0-\varepsilon V\,,
\hskip 1cm 
H_{\varepsilon}=H_0+\varepsilon V\,, 
\]
with $L_0$, $H_0$ the Lagrangian and Hamiltonian of the 2-centre problem
\[
L_0=\frac{\dot{x}^2+\dot{y}^2}{2} - W\,,
\hskip 1cm
H_0=\frac{p_x^2+p_y^2}{2}+W\ .
\]

For negative values of the energy $H_{\varepsilon}=E<0$, with $E
\rightarrow 0$, we prove the existence of chaotic motions passing
arbitrarily close to the perturbing centre $C$.  This is made by the
use of the shadowing result proved by Bolotin and MacKay in
\cite{BM1}.

For $E>0$ non-integrability has been established by Bolotin (see
\cite{Bol1}): he proves that for the $n$-centre problem on the plane
$\R^2$, with $n>2$, it does not exist an analytic integral of motion
which is non-constant on the energy shell $H^{-1}(E)$, with $E>0$.
In~\cite{Bol2} the same author extends this result to a wider class of
Lagrangian systems defined on any 2-dimensional manifold $M$, with $n$
Newtonian singularities on $M$: he shows non-integrability when $n$ is
greater than two times the Euler characteristic of the manifold,
$n>2\chi(M)$, and the energy is over a suitably defined threshold,
$E>E_{th}$.  In particular, we have analytic non-integrability for the
restricted circular many-body problem, in which a particle moves in a
rotating plane, under the action of the gravitational attraction of
$n$ centres fixed on this plane, when $n>2$.  Nevertheless, this
generalisation does not add any additional information about the
$n$-centre problem on a fixed plane, in which case it reduces to the
result given in~\cite{Bol1}.

More recent outcomes exist in literature, valid for positive energy
values $E>0$: see~\cite{BN1}, \cite{KT}.  In~\cite{BN1} the authors
study the $n$-centre problem in $\R^3$, showing that if $n\geq3$ and
$E \geq 0$, then the topological entropy is positive.  In~\cite{KT},
it is proved that no analytic independent integral exist for the
$n$-centre problem in the space, if $n \geq 3$ and the energy is
greater than some threshold $E>E_{th}$.  Moreover, for both the planar
and the spatial problem, smooth integrals are found on energy levels
with $E>E_{th}$, where $E_{th}=0$ in the planar case.

The case $E<0$ has been investigated in~\cite{BN2}. The authors study
the restricted 3-centre problem on the plane, when the third centre is
very far from the other two and consider small negative energies $E$,
in the limit $E\rightarrow 0$. Then they have a two-parameter
perturbation of the 2-centre problem on the zero-energy level. They
succeed in applying the Poincar\'e-Melnikov theory, thus proving the
existence of chaotic motions.

We too study the case of small negative energies, but our point of
view is different, in that the position of the third centre does not
go to infinity.
This allows us to study the orbits which undergo close encounters with
the perturbing centre $C$.

As in~\cite{BN2} the unperturbed system is the 2-centre problem with
Lagrangian $L_0$. After the classical regularisation of singularities,
obtained by the use of elliptic coordinates $(\xi, \phi) \in \R \times
S^1$ and a reparametrisation of time, the system separates: we have
the motion of the standard pendulum for the variable $\phi$ and a
two-well potential for $\xi$, with a hyperbolic equilibrium for $\xi =
0$.
We are interested in the motions which go very close to the centre
$C$, then we consider trajectories of the unperturbed problem that
pass through $C$.
The perturbing term of the Hamiltonian is not $\varepsilon$-small in
the vicinity of a collision with $C$ and this results in a lost of
control on the stable and unstable manifolds, so that a perturbative
study of their intersection is not possible.

Anyway, we will see that for $\varepsilon$ small enough a hyperbolic
invariant set with chaotic dynamics exists, for a dense set of
possible positions of the third centre in $\R^2$.  Moreover, the
obtained trajectories shadow chains of collision orbits through
the third centre $C$.
  
We will apply a general result of Bolotin and MacKay (see \cite{BM1}),
valid for Lagrangian systems with Newtonian singularities on a
$d$-dimensional Riemannian manifold, with $d=2,3$: they show that if
the singular part of the Lagrangian is a perturbing term, then for
each chain formed by arcs of the unperturbed problem which start and
end at a singular point (\emph{collision arcs}), there is a unique
orbit with the same energy which shadows it.
Indeed, after local regularisation of the perturbing singularities,
the corresponding equilibria are hyperbolic and, by the use of the
strong $\lambda$-lemma, shadowing arcs are obtained locally in small
neighbourhoods of the singularities. It is then proved
that these arcs can be joined to gain the entire shadowing orbits.
%
In fact, the latter correspond to nondegenerate critical points of a formal
functional defined on small spherical neighbourhoods (circular in the
2-dimensional case) of the singularities. This functional can be seen
as the energy functional of a Frenkel-Kontorova model. There is a
correspondence between the equilibrium states of Frenkel-Kontorova
models and the orbits of symplectic twist maps, and in this
correspondence nondegeneracy of the equilibrium states of the energy
functional corresponds to hyperbolicity, as shown in \cite{AMB}.
With this approach the existence and hyperbolicity of the shadowing
orbits is proved and, by proper estimates on the mixed second
derivatives of the formal functional, it is also derived that the
Lyapunov exponents of the corresponding Poincar\'e map are of order
$\log \varepsilon^{-1}$(see \cite{BM2}).

The result of \cite{BM1} has been applied by the two authors to 
the planar circular restricted problem of three bodies, in which a
massless particle moves under the gravitational attraction of the
other two bodies, the primaries, and the latters are supposed on a
circular orbit about their centre of mass.  The second primary is
supposed to have small mass $\varepsilon$.  The result is the
existence of periodic and chaotic orbits which undergo consecutive
close encounters with the smaller primary and which approach, as
$\varepsilon \rightarrow 0$, arcs of Kepler ellipses around the first
primary, starting and ending at collision with the second.

A similar existence result has been obtained also in \cite{FNS1} by a
completely different method. A direct study of the orbits which have
close encounters with the small primary is carried out and the proof
in this case is constructive. Indeed, an approximation of the
\emph{first return map}, defined on a region of the phase space whose
projection is a small circle around the second primary, is explicitly
computed.  It is shown that the first return map is \emph{horseshoe}
like and it allows to conclude about the existence of orbits with
consecutive infinite close approaches with the small primary.  A
complete numerical study of these orbits is carried out in
\cite{FNS2}.

As previously outlined, in this paper we follow the variational
approach of Bolotin and MacKay applied to the 3-centre problem: in
this manner we don't get an horseshoe map, but we still have a
symbolic dynamics on the set of quasi-collision orbits found.

In our case the unperturbed system is simply the 2-centre problem with
Lagrangian $L_0$ and the perturbing term is given by the potential due
to the third centre $C$. Fixed a small negative value of the energy
$E<0$, by the classical regularisation of the 2-centre problem the
system separates and we can easily find periodic orbits with energy
$E$ passing through $C$. We prove by the implicit function theorem
that for almost all the possible positions of the perturbing centre
$C$ on the plane, these orbits do not collide with the primaries
$C_1,C_2$, if $|E|$ is chosen sufficiently small.  Then they are
periodic orbits of the not-regularised 2-centre problem and we can use
them to define a finite set of collision arcs, starting and ending at
$C$, and take as collision chains the infinite sequences formed with
the chosen arcs.  We can verify that the considered orbits of the
unperturbed problem are nondegenerate (see Section~\ref{bolotin} for
the definition) and then the theorem of Bolotin and MacKay can be
applied to obtain a hyperbolic invariant set formed by the orbits that
shadow the collision chains.
Finally, taking as alphabet the finite set of collision arcs chosen, a
symbolic dynamics is naturally defined on the set of shadowing orbits
by considering the shift on the collision chains.  We can also get
easily the positivity of entropy.
 
In Section~\ref{bolotin}, we will recall the results of \cite{BM1} and
\cite{BM2} for the case at hand , starting from some basic
definitions, and state our main result.  In
Section~\ref{regular_periodic} we will perform the classical
regularisation of the 2-centre problem and compute the periods of the
resulting separated problem. This will allow us to get infinite
classes of periodic orbits with fixed energy $E$, passing through the
centre $C$ (Section~\ref{collision}): this is the core of the
paper. The delicate point here is to show that the periodic orbits of
the regularised problem previously found do not meet the primaries for
any small enough value of the energy.  In
Section~\ref{degen_dir_change} we will verify that these trajectories
satisfy the conditions which allow to apply the results of \cite{BM1},
\cite{BM2} and conclude our proof.

\section{Basic definitions and theorems.}
\label{bolotin}
Consider the unperturbed system with Lagrangian $L_0$.  A solution
$\gamma:[0,T]\rightarrow M$ of fixed energy $E$ for this system will
be called a \emph{collision arc} if $\gamma(0)=\gamma(T)=C$, and
$\gamma(t)\neq C$ for any $t\in (0,T)$. In particular, the latter
condition means that there are no \emph{early} collisions.

Fix an energy value $E<0$, such that $C$ belongs to the set
$D=\{(x,y)\in M|\ W(x,y)<E\}$.  Denote by $\Omega$ the set of
$W^{1,2}$ curves lying in $D$, starting and ending at collision in
$C$.  A collision arc $\gamma $ of energy $E$ is a critical point of
the Maupertuis-Jacobi functional $J_E$ defined on $\Omega$ by
\[
J_E(\gamma)=\int_0^Tg_E(\gamma(t),\dot\gamma(t))dt\,, 
\]
with 
\[
g_E(x,y, \dot{x},\dot{y})=\sqrt{2(E-W(x,y))(\dot{x}^2 + \dot{y}^2)}\ .
\]
The arc $\gamma$ is said to be \emph{nondegenerate} if it is a
nondegenerate critical point of the functional $J_E$ on $\Omega$.

This definition of nondegeneracy is the most natural, but it is quite
complicated to be used for verifications in concrete examples. There
are at least four equivalent definitions of nondegeneracy, for which
we refer the reader to \cite{BM1}.  In particular, a sufficient
condition for nondegeneracy is the following.  Denote the general
solution of the system $(L_0)$ by $q(t)=f(q_0,v_0,t)$, where $q_0 \in
M, v_0 \in TM_{q_0}$ are the initial position and velocity.  Then a
collision arc $\gamma$ with energy $E$ corresponds to a solution
$(v_0,T)$ of the system of three equations
\[
f(C,v_0,T)=C\,, \quad H_0(C,v_0)=E\ .
\]
If the Jacobian of the system at this solution is nonzero, then
$\gamma$ is nondegenerate.  Actually this is the characterisation of
nondegeneracy which will be used in Section~\ref{degen_dir_change}.

Suppose that the system $(L_0)$ has a finite set of nondegenerate
collision arcs through the centre $C$, $\gamma_k: [0,T_k] \rightarrow
D$, $k \in K$, with the same energy $E$, where $K$ denote a finite set
of labels.  A sequence $(\gamma_{k_i})_{i \in \Z}$, $k_i \in K$, is
called a \emph{collision chain} if it satisfies the condition of
\emph{direction change}:
\[
\dot{\gamma}_{k_i}(T_{k_i}) \neq \pm \dot{\gamma}_{k_{i+1}}(0)\,, 
\textrm{ for any } i \in \Z \ .
\]
Collision chains correspond to paths in the graph $\Gamma$ with the
set of vertices $K$ and the set of edges
\begin{equation}
\label{graph}
\Gamma=\{(k,k') \in K^2| \ 
\dot{\gamma}_{k}(T_{k}) \neq \pm \dot{\gamma}_{k'}(0)\}\ .
\end{equation}
We are going to use the following results:
\begin{theorem}[Bolotin-MacKay, 2000]
\label{bol-mack}
Given a finite set $K$ of nondegenerate collision arcs with the same
energy $E$, there exists $\varepsilon_0>0$ such that for all
$\varepsilon \in (0,\varepsilon_0]$ and any collision chain
  $(\gamma_{k_i})_{i \in \Z}$, $k_i \in K$, there exists a unique (up
  to a time shift) trajectory $\gamma: \R \rightarrow D\setminus
  \{C\}$ of energy $E$ of system $(L_{\varepsilon})$, which shadows
  the chain $(\gamma_{k_i})_{i \in \Z}$ within order $\varepsilon$.
  More precisely, there exist constants $B, B' > 0$, independent of
  $\varepsilon$ and the collision chain, and a sequence of times
  $(t_i)_{i\in \Z}$, such that $|t_{i+1}-t_i-T_{k_i}|\le
  B\varepsilon$, $dist(\gamma(t),\gamma_{k_i}([0, T_{k_i}])) \le
  B\varepsilon$, for $t_i\le t \le t_{i+1}$, and $dist(\gamma(t),C)\ge
  B'\varepsilon$.
\end{theorem}
From Theorem~\ref{bol-mack} it follows that there is an invariant
subset $\Lambda_{\varepsilon}$ on the energy shell
$\{H_{\varepsilon}=E\}$ on which the system $(L_{\varepsilon})$ is a
suspension of a subshift of finite type.  The subshift is given by the
shift on the set of paths of the graph $\Gamma$, that is on the set of
all the possible collision chains, and the set $\Lambda_{\varepsilon}$
is formed by the orbits which shadow them.

The important fact about the invariant set $\Lambda_{\varepsilon}$ is
that it is uniformly hyperbolic.

\begin{theorem}[Bolotin-MacKay, 2006]
  \label{bol-mack_hyperb}
There exists a cross-section $N \subset \{H_{\varepsilon} = E\}$, such
that the corresponding invariant set
$M_{\varepsilon}=\Lambda_{\varepsilon}\cap N$ of the Poincar\'e map is
uniformly hyperbolic with Lyapunov exponents of order $\log
\varepsilon^{-1}$.
  
In particular, the set $\Lambda_{\varepsilon}$ is uniformly hyperbolic
as a suspension of a hyperbolic invariant set with bounded transition
times.
\end{theorem}
The reader can see~\cite{BM2} for the construction of the
cross-section $N$ and the definition of the Poincar\'e map: we omit
these technical details here, because it would require some machinery,
which is not necessary for the sequel.
  
We will show that the assumptions of Theorem~\ref{bol-mack} are
all satisfied for the case at hand, for almost all the possible
positions of the third centre $C$ in $M$.  More precisely, we will
prove that, fixed any of these admissible positions of $C$, for any
small enough negative energy value, there exist non-degenerate
collision arcs, with which we can construct collision chains.  In
particular, our collision arcs are pieces of periodic orbits of the
$2$-centre problem $(L_0)$, which pass through the centre $C$.  We
will then derive the following
\begin{theorem}
\label{central_result}
Let $I \subset \Q^+$ be a finite set of positive rationals.  There
exists a dense open subset $X_I \subset M$ of possible positions for
the third centre $C$, such that, fixed $C \in X_I$, the following is
true.  There is a small value $E_0>0$, depending on $C$ and $I$, such
that, fixed an energy value $E \in (-E_0,0)$, we have that there is
$\varepsilon_0>0$, such that for any $\varepsilon \in
(0,\varepsilon_0)$ and any sequence $(q_{k})_{k \in \Z}, q_{k}\in I$,
there exists a trajectory of the planar restricted 3-centre problem
$(L_{\varepsilon})$ on the energy shell $\{H_{\varepsilon}=E\}$, which
avoids collision with the third centre $C$ by order $\varepsilon$ and
is within order $\varepsilon$ a concatenation of pieces of periodic
orbits for the planar restricted 2-centre problem $(L_0)$, passing
through $C$ and of classes $q_k, k \in \Z$ (see
Subsection~\ref{per_orb} for notation).
  
The resulting invariant set formed by these orbits is uniformly
hyperbolic.
\end{theorem}
In particular, as it follows from Theorem~\ref{bol-mack_hyperb}, fixed
a small enough energy value $E<0$ and $\varepsilon >0$, there is a
cross section in the energy shell $\{H_{\varepsilon}=E\}$, such that
the associated Poincar\'e map has a chaotic invariant set with
Lyapunov exponents of order $\log \varepsilon^{-1}$ and it contains
infinitely many periodic orbits, corresponding to periodic collision
chains.
The topological entropy of the Poincar\'e map is
$O(\varepsilon)$-close to that of the topological Markov chain
associated to the graph $\Gamma$, defined by (\ref{graph}).  It's easy
to verify that the finite set of collision arcs that we construct in
the proof of Theorem~\ref{central_result} determine positive
topological entropy (see Appendix~\ref{app:entropy}).

Theorem~\ref{central_result} is still true if we substitute in the
statement the set $X_I$ with a set $X$, which is dense in $M$ and is
independent of the set of rationals $I$.  In this case, we do not know
if the set $X$ is open or not, the only thing that we can say about it
is that it is dense (see Remark~\ref{XX_I}).

\section{Periodic orbits for the regularised problem.}
\label{regular_periodic}
In order to apply Theorem~\ref{bol-mack}, we must find collision arcs
with fixed energy $E<0$ for the unperturbed system $(L_0)$.  A natural
choice is to look for periodic orbits through the third centre $C$.
As a first step, we recall the classical regularisation of
singularities of Euler.  A deep analysis of the 2-centre problem was
made by Charlier in the first volume of \cite{Ch}.  Following his
ideas we get the existence of infinite classes of periodic orbits for
the separated problem obtained after regularisation\footnote{A
  classification of the periodic orbits of the 2-centre problem based
  on the variation of the energy parameters can be found in \cite{DY}
  and \cite{DYB}.  We will not need this classification: according to
  it our orbits are all of the same kind, because of the constraints
  that we impose on the parameters.}.

\subsection{Regularisation}
We introduce the elliptic coordinates, defined by the map
$x+iy=\cosh(\xi+i\phi)$ from the cylinder $\R \times S^1$ to $\R^2$.
This transformation has two ramification points at the two primaries
$C_1=(0,0),C_2=(0,\pi)$.  In elliptic coordinates the Lagrangian
$L_{\varepsilon}$ becomes
\[
L_{\varepsilon}=\frac{\dot{\xi}^2+\dot{\phi}^2}{2}(\cosh^2\xi-\cos^2\phi)
-W(\xi,\phi) -\varepsilon V(\xi,\phi)\,,
\] 
where the potentials $W,V$ have the form
\begin{eqnarray*}
W(\xi,\phi)&=&-\frac{2 a \cosh \xi}{\cosh^2\xi-\cos^2\phi} \,,
\\ 
V(\xi,\phi)&=&-\frac{1}{\sqrt{\cosh^2\xi-\sin^2\phi+(x_0^2+y_0^2)
    -2(x_0\cosh\xi\cos\phi+y_0\sinh\xi\sin\phi)}} \ .
\end{eqnarray*}
Consider the problem on the energy level set $\{H_{\varepsilon}=E\}$.
The Hamiltonian in elliptic coordinates is
\[
H_{\varepsilon}-E = 
\frac{1}{\cosh^2 \xi - \cos^2\phi}\mathcal{H}_{\varepsilon}\,,
\]
with
\[
\mathcal{H}_{\varepsilon} = \frac{{p_\xi}^2  + {p_\phi}^2}{2} -
2a \cosh \xi-\left[E- \varepsilon V(\xi,\phi)\right]
(\cosh^2 \xi - \cos^2\phi)\,,
\]
where the symbols $p_{\xi},p_{\phi}$ denote the conjugate momenta.

The orbits of the problem with Hamiltonian $(H_{\varepsilon})$ on the
energy level $\{H_{\varepsilon}=E\}$ are, up to time parametrisation,
orbits of the system with Hamiltonian $\mathcal{H}_{\varepsilon}$ on
the energy level $\{\mathcal{H}_{\varepsilon}=0\}$. The regularised
Hamiltonian $\mathcal{H}_{\varepsilon}$ has no singularities at
$C_1,C_2$: this means that an orbit for $\mathcal{H}_{\varepsilon}$ is
an orbit for $H_{\varepsilon}$ only if it does not pass through the
primaries.  The new time parameter $\tau$ is given by
\begin{equation}
  \label{newtime}
  \tau=\int_0^t\frac{1}{\cosh^2\xi(s)-\cos^2\phi(s)}ds\,,
\end{equation}
and this formula allows us to pass from a solution for the system
$(\mathcal{H}_{\varepsilon})$, on the zero energy level, to a solution
for $(H_{\varepsilon})$ with energy $E$.

The Lagrangian corresponding to $\mathcal{H}_{\varepsilon}$ is
\[
\mathcal{L}_\varepsilon = \frac{(\xi')^2  + (\phi')^2}{2} +
2 a \cosh \xi  + \left[E-\varepsilon V(\xi,\phi)\right]
(\cosh^2 \xi - \cos^2\phi)\,,
\]
where the prime sign denote derivation with respect to the new time
parameter: $\xi'=d\xi/d\tau$, $\phi'=d\phi/d\tau$.
\subsection{The separated problem.}
\begin{figure}[h]
  \begin{minipage}[b]{0.49\textwidth} 
   \centering
    \includegraphics[width=\textwidth]{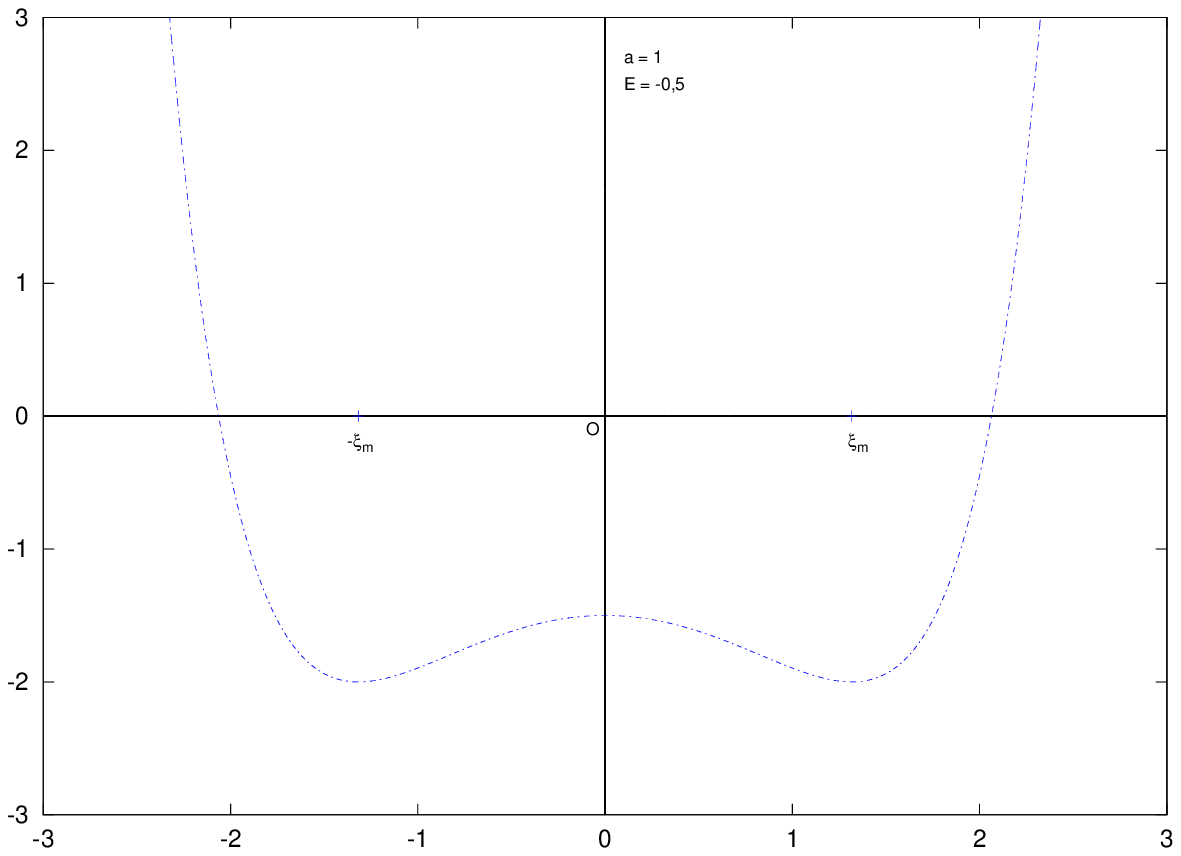}
    \caption{\small{Potential energy for the variable $\xi$ when $a=1$ and 
        $E=-0.5\,$.}}
    \label{fig:potenxi}
  \end{minipage}
  \hspace{2mm} 
  \begin{minipage}[b]{0.49\textwidth} 
    \centering
    \includegraphics[width=\textwidth]{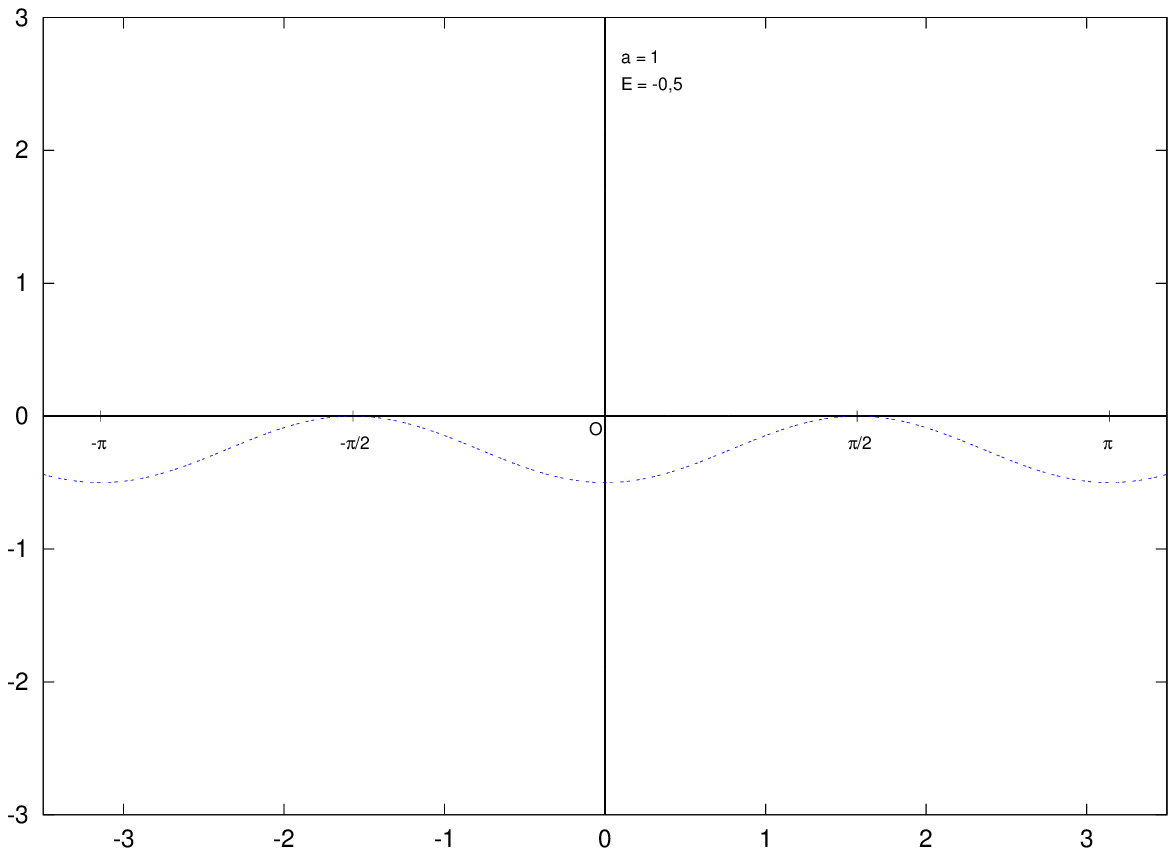}
    \caption{\small{Potential energy for the variable $\phi$ when $a=1$ 
        and $E=-0.5\,$.}}
    \label{fig:potenphi}
  \end{minipage}
\end{figure}

We have to find orbits of the 2-centre problem $(L_0)$ with fixed
energy $E<0$, then we put $\varepsilon=0$ and study the regularised
system on the energy level $\{\mathcal{H}_0=0\}$.  The Lagrangian is
\[
\mathcal{L}_0=\frac{(\xi')^2  + (\phi')^2}{2} +
2 a \cosh \xi  + E(\cosh^2 \xi-\cos^2\phi) \ .
\]
The system separates and we have the two one-dimensional problems
\begin{equation}
  \label{sys_reg}
  \left\{
  \begin{array}{rcl}
    \frac{(\xi')^2  }{2} -
    2a \cosh \xi  + |E|\cosh^2 \xi &= &  
    -E_1
    \\
    \frac{ (\phi')^2}{2} 
    - |E| \cos^2\phi&= & E_1
  \end{array}
  \right.
\end{equation}
where we have taken into account that $E<0$. If we assume $|E|<a$,
then the potential in the variable $\xi$ has a local maximum when
$\xi=0$, with negative maximum value $|E|-2a$, and two minima
$\pm\xi_{m}$, defined by $\cosh(\pm \xi_{m})=a/|E|$, where the
potential takes the value $-a^2/|E|$. Furthermore, it goes to infinity
for $\xi \rightarrow \pm \infty$.  Finally, if we choose
$-E_1>|E|-2a$, i.e. we consider a level over the separatrix, the
motion in $\xi$ is periodic with two inversion points at
$\xi_-$ and $\xi_+$, defined by the relation:.
\[
\cosh(\xi_{\pm})=
-\frac{a}{E}\left(1+\sqrt{1+\frac{EE_1}{a^2}}\right)
\ .
\]  
The motion in $\phi$ is simply the motion of the standard pendulum.
Then, in the domain $E_1>0$, we have rotational closed orbits for
$\phi$ and there exist action-angle variables.  The graphs of the
potential energies for $\xi$ and $\phi$ separately are plotted in
Figures~\ref{fig:potenxi} and \ref{fig:potenphi}, when $a=1$ and
$E=-0.5$ .

After these considerations, we make the following assumptions:
\begin{equation}
  \label{hyp_E}
  |E|<a\,, \quad E_1>0\,, \quad |E|+E_1<2a\ .
\end{equation} 
For future convenience, we scale the energy parameter $E_1$ and define
\[
A_1:=E_1/2a\,, \quad  \beta:= |E|/E_1 \ .
\]
In this manner the conditions~(\ref{hyp_E}) becomes
\begin{equation}
  \label{hyp_A}
  A_1,\beta>0\,, \quad 2\beta A_1 < 1\,, \quad A_1<\frac{1}{1+\beta}\,,
\end{equation}
and the regularised system~(\ref{sys_reg}) takes the form
\begin{equation}
  \label{sys_A}
  \left\{
  \begin{array}{rcl}
    \frac{(\xi')^2  }{4a}&=& 
    \cosh \xi  - \beta A_1\cosh^2 \xi   
    - A_1
    \\
    \frac{ (\phi')^2}{4a} 
    &=&  \beta A_1 \cos^2\phi +A_1 
  \end{array}
  \right.
\end{equation}

With the assumptions~(\ref{hyp_A}) on $\beta,A_1$, both the
one-dimensional motions are periodic, with periods $T_1,T_2$
respectively for $\xi,\phi$, which depend only on the values of
$\beta,A_1$. If the periods $T_1,T_2$ have rational ratio $T_1/T_2 \in
\Q$, then the corresponding orbit on the cylinder $\R\times S^1$ is
periodic. To find a periodic orbit for the system $(\mathcal{L}_0)$,
corresponding to a fixed value of the energy parameter $\beta$, we
have to show that there is at least a value of $A_1$ for which
$T_1,T_2$ have rational ratio.  Before doing that we must compute the
analytical expressions of the periods.
\begin{lemma}
\label{periods}
Consider the system~(\ref{sys_A}), with the
assumptions~(\ref{hyp_A}). Then the one-dimensional motions are both
periodic with periods $T_1,T_2$, for the coordinates $\xi,\phi$
respectively, given by
\begin{eqnarray}
    T_1=\frac{2\sqrt{2a^{-1}}}{\sqrt[4]{1-4\beta A_1^2}}\K(\kappa_1)
    \,, &\quad&
    T_2=\frac{2\sqrt{a^{-1}}}{\sqrt{A_1(1+\beta)}}\K(\kappa_2) \,,
\end{eqnarray}
where
\[
\kappa_1^2=\frac{A_1(1-\beta)+\sqrt{1-4\beta A_1^2}}{2\sqrt{1-4\beta
    A_1^2}} \,, \quad \kappa_2^2=\frac{\beta}{1+\beta} \,, \quad
\kappa_1,\kappa_2 > 0 \,,
\] 
and $\K(\kappa)$ is the elliptic integral of the first type:
\[
\K(\kappa)=\cn^{-1}(0,\kappa)=\int_0^1\frac{dv}{\sqrt{(1-v^2)(1-\kappa^2v^2)}}
\ .
\]
\end{lemma}
\begin{proof} 
It is a straightforward computation, starting from the integral
expressions of $\tau(\xi),\tau(\phi)$. 
\qed
\end{proof}

\begin{remark}
  \label{lim-T1T2}
  Observe that $\lim_{\kappa \rightarrow 1^-}\K(\kappa)=+\infty$ and 
  \[
  \kappa_1=1 \iff \left\{
  \begin{array}{rcl}
    \beta &<& 1
    \\
    A_1&=&\frac{1}{1+\beta}
  \end{array}
  \right.\ .
  \]
  In particular, the second condition means that the motion of $\xi$ takes 
  place on the separatrix energy level. Then, if $\beta < 1$ 
  the period $T_1$ goes to infinity as $A_1 \rightarrow \frac{1}{1+\beta}$:
  \[
  \lim_{A_1 \rightarrow \frac{1}{1+\beta}}T_1= +\infty\ .
  \]
  Correspondingly, for $T_2$ we have:
  \[
  \lim_{A_1\rightarrow 0}T_2=+\infty \ .
  \]
\end{remark}

We are interested to the limit $E \rightarrow 0$, 
then we can suppose that the  parameter $E_1$ is greater than $|E|$.
It corresponds to make the hypothesis that $\beta<1$.
Our definitive assumptions are:  
\[
\beta \in (0,1)\,,
\quad
0<A_1<\frac{1}{1+\beta}\ .
\] 
\subsection{Periodic orbits.}
\label{per_orb}
The expressions of the periods $T_1,T_2$ given in Lemma~\ref{periods}
and the Remark~\ref{lim-T1T2} give the limits
\begin{equation}
\label{limits}
\begin{array}{rcl}
\lim_{A_1\rightarrow\frac{1}{1+\beta}}T_1= +\infty \,, 
&\quad&
\lim_{A_1 \rightarrow 0}T_2=+\infty \,, \nonumber
\\ 
\lim_{A_1\rightarrow 0}T_1=\frac{4}{\sqrt{2a}}\K(\frac{1}{\sqrt{2}}) \,,
&\quad &
\lim_{A_1\rightarrow\frac{1}{1+\beta}}T_2=\frac{2}{\sqrt{a}}
\K(\sqrt{\frac{\beta}{1+\beta}}) \ .
\end{array}
\end{equation}
Moreover, from the definition of $\K(\kappa)$, we easily conclude that, fixed 
$\beta \in (0,1)$, $T_2$ is a strictly decreasing function of 
$A_1 \in (0,\frac{1}{1+\beta})$, while $T_1$ is strictly increasing. 
Indeed, $K(\kappa)$ is a strictly increasing function of $\kappa^2$,
$\kappa_2$ does not depend on $A_1$ and $\kappa_1^2$ has positive 
derivative with respect to $A_1$ given by
\[
\frac{\partial \kappa_1^2}{\partial A_1} = 
\frac{1-\beta}{2}(1-4\beta A_1^2)^{-\frac{3}{2}} >0\ .
\]   
By these 
observations, the following is shown.
\begin{proposition}
  \label{periodic_orbits}
  Let $\beta \in (0,1)$ be fixed. For any positive rational $q \in \Q^+$, there 
  exists a unique value $\hat{A}_1(\beta,q) \in (0,\frac{1}{1+\beta})$ for 
  $A_1$, such that
  \begin{equation}
    \label{qT}
    qT_1(\beta,\hat{A}_1)=T_2(\beta,\hat{A}_1)\ . 
  \end{equation}
  In particular, the system~(\ref{sys_A}) has a periodic solution in 
  correspondence of the value $\hat{A}_1$, with energy 
  $E=-2 a \beta \hat{A}_1$. 
\qed
\end{proposition}

We have seen that for any $\beta \in (0,1)$ and any positive rational
$q \in \Q^+$, there is a periodic orbit for the regularised system
$(\mathcal{L}_0)$ with energy given by the relation $E=-2a \beta
\hat{A}_1$.  We can classify these periodic orbits, identifying each
\emph{class} with the rational number $q$. Then for any fixed value of
the parameter $\beta \in (0,1)$, we have exactly one value of
$\hat{A}_1$ for each class $q \in \Q^+$. Orbits of different classes
do not have the same energy $E$; more precisely we have
\begin{proposition}
  \label{A1_respect_q}
  Let $\beta \in (0,1)$ be fixed. The function 
  $\hat{A}_1(\beta, \cdot): \Q^+ \rightarrow (0,\frac{1}{1+\beta})$, 
  defined by the equality (\ref{qT}), is strictly decreasing. 
  Moreover, there exist the limits
  \[
  \lim_{q \rightarrow 0^+}\hat{A}_1(\beta,q)=\frac{1}{1+\beta}
  \,, \hskip 0.5cm
  \lim_{q \rightarrow +\infty}\hat{A}_1(\beta,q)=0
  \ .
  \]
\end{proposition}
\begin{proof}
  If $q>q'$ then $qT_1(\beta,A_1)>q'T_1(\beta,A_1)$ and we have 
  \[
  T_2(\beta, \hat{A}_1(\beta,q))=qT_1(\beta, \hat{A}_1(\beta,q))
  >q'T_1(\beta,\hat{A}_1(\beta,q)))\ .
  \] 
  $T_2$ is strictly decreasing with respect to $A_1$, while $T_1$ is
  strictly increasing, then $\hat{A}_1(\beta,q)<\hat{A}_1(\beta,q')$.
  
  The monotony of $\hat{A}_1(\beta,\cdot)$ assures the existence of the limits
  \[
  L_1=\lim_{q \rightarrow 0^+}\hat{A}_1(\beta,q)\,, \quad
  L_2=\lim_{q \rightarrow +\infty}\hat{A}_1(\beta,q)\ .
  \]
  Clearly $L_1,L_2 \in \left[0,\frac{1}{1+\beta}\right]$ and from the knowledge 
  of the limits (\ref{limits}), we easily obtain the desired values for 
  $L_1,L_2$.
\qed
\end{proof}

It follows that periodic orbits with many ``loops'' in $\xi$ and few
in the variable $\phi$ tend to the separatrix level for $\phi$;
viceversa, orbits with many loops in $\phi$ tend to the separatrix
level for $\xi$.  In other words, if we increase only for a single
variable the number of loops before the orbit closes, we will obtain a
limit orbit which does not close anymore in finite time.  For example,
take $q=m/n$, with $m,n \in \N$. Increasing the number of loops in
$\xi$ corresponds to make $m$ larger and consequently $\hat{A}_1$
smaller.  The periodic orbit increases the number of ``oscillations''
in $\xi$, while making the same number of revolutions in the variable
$\phi$.  As $m$ goes to infinity we have that the corresponding orbit
makes an infinite number of times the same trajectory in $\xi$,
tending to close, but without being able to reach the limit values
$\phi=\pm \pi/2$ in a finite time interval, in the future and in the
past respectively: in particular, it does not complete even a single
revolution for $\phi$.  Furthermore, the energy $E=-2a \beta
\hat{A}_1$ tends to zero.

We conclude that to form easily a finite set of collision arcs with
the same energy, we should fix $q \in \Q^+$ and look for collision
arcs only in the set of periodic orbits of the same class $q$.
 
\section{Construction of collision arcs.}
\label{collision}
In the previous section we have found infinite classes of periodic
orbits for the regularised problem $\mathcal{L}_0$. We would like to
use them to construct collision arcs.
 
The next step is then to show that, among the periodic orbits of the
regularised system $(\mathcal{L}_0)$, there is at least one which
passes through the third centre $C$.  Actually this is true if the
parameter $\beta$ is sufficiently small.  We must also verify that the
obtained orbits are solutions of the not regularised problem $(L_0)$,
that is they do not pass through the primaries: this is the most
delicate point of the proof of Theorem~\ref{central_result}.
Moreover, in Subsection~\ref{early_coll} we will face the problem of
\emph{early collisions}.

\subsection{Periodic orbits through the third centre.}
\label{coll_C}
Let $(\xi_0,\phi_0) \in \R \times S^1 \setminus \{(0,0),(0,\pi)\}$ be
fixed elliptic coordinates for the position of the third centre $C$.
Then, among the orbits corresponding to the value
$\hat{A}_1(\beta,q)$, surely there is one which pass through the
centre $C$, if $\xi_0 \in (\xi_-(\beta,\hat{A}_1),
\xi_+(\beta,\hat{A}_1))$, where $\xi_{\pm}$ are the inversion points.
Note that in Cartesian coordinates this corresponds to say that the
centre $C$ lies in the region internal to the ellipse defined by the
equation $\xi=\xi_+$.
By construction, for each $\beta \in (0,1)$, we have $\hat{A}_1 \in
(0, \frac{1}{1+\beta})$, then $\lim_{\beta \rightarrow 0}\beta
\hat{A}_1=0$ and
\[
\lim_{\beta \rightarrow 0} \cosh(\xi_{\pm})= 
\lim_{\beta \rightarrow 0} \frac{1+\sqrt{1-4\beta\hat{A}_1^2}}{2\beta\hat{A}_1}=
+\infty
\ .
\]
Thus we conclude that
\begin{proposition}
  \label{collisionC}
  Fixed a class $q \in \Q^+$,
  there exists $\beta_0>0$, such that for any $\beta \in (0,\beta_0)$
  there is a  periodic orbit of system $(\mathcal{L}_0)$, 
  associated with the value $\hat{A}_1(\beta,q)$, which passes through $C$, 
  and the coordinate $\xi_0$ is not an 
  inversion point of the corresponding one-dimensional motion in $\xi$. 
\qed
\end{proposition}

Note that the periodic orbits associated with the same value of
$\hat{A}_1$ differ only for the sign of the velocities
$\xi'_0,\phi'_0$ at the centre $C$.  Thus we have exactly two orbits
on the configuration space $\R \times S^1$: if one has velocity
$(\xi'_0,\phi'_0)$ at $C$, the other has velocity
$(-\xi'_0,\phi'_0)$. The remaining two possibilities give the same
orbits, but with the opposite direction of motion. Moreover
$\xi'_0\neq 0$, because $\xi_0$ is not an inversion point. Then the
two trajectories corresponding to $\hat{A}_1$ meet transversely at $C$
on the cylinder $\R \times S^1$.

We remark that there is the possibility that the two orbits coincide:
it can happen when the trajectory has an autointersection at
$(\xi_0,\phi_0)$ before closing. This is a case of \emph{early
  collision} and it will be treated in
Proposition~\ref{ellip_early_coll}.

The obtained solutions are not yet the collision arcs that we
desire. In fact, they are orbits for the regularised system
$(\mathcal{L}_0)$: to be orbits of the 2-centre problem with
Lagrangian $L_0$, it's enough they do not pass through the primaries
$C_1,C_2$.  This is a delicate problem and to face it we will need
a general result about the regularity of $\hat{A}_1$ as function
of $\beta$, in a neighbourhood of $\beta = 0$.

\subsection{Avoiding collision with the primaries.}
\label{collision_primaries}
In this subsection we will obtain the central result of the paper: we
will show that for almost all the possible positions of the third
centre in $\R \times S^1$, the periodic orbits through $C$
corresponding to $\hat{A}_1$ don't collide with the primaries for any
sufficiently small value of $\beta$.  It means that they are solutions
of the not-regularised system $(L_0)$ and allows us to proceed with
the final verifications, in order to apply Theorem~\ref{bol-mack}.

First of all we would like to recall a useful property which
characterises the periodic orbits of the regularised 2-centre problem
passing through one of the primaries. Note that periodic orbits
through the primaries always exist because a primary has $\xi=0$.

\begin{proposition}
\label{coll_primaries}
Let $\beta \in (0,1)$ arbitrarily fixed.  Given $q \in \Q^+$, let $m,n
\in \N$ such that $q=m/n$ and $(m,n)=1$.  Let $\gamma$ be a periodic
orbit of the system $(\mathcal{L}_0)$ associated with
$\hat{A}_1(\beta,q)$ and suppose that it passes through one of the
centres $C_1,C_2$, which have elliptic coordinates $(0,0), (0,\pi)$
respectively.
  
If $n$ is odd, then the orbit goes through both the primaries in a
period, and the collisions happens at a time distance of half the
period from each other.
  
If $n$ is even, then the orbit passes through only one of the
primaries and it happens two times in a period, at a time distance of
half the period.  In the configuration space $\R \times S^1$ the orbit
has a transverse self-intersection at the position of the centre.
\end{proposition}
\begin{proof}
The system $(\mathcal{L}_0)$ has the form (\ref{sys_A}).  Without loss
of generality we can suppose that the orbit
$\gamma(\tau)=(\xi(\tau),\phi(\tau))$ passes through one of the
centres $C_1,C_2$ at time $\tau =0$.  The orbit $\gamma$ collides with
a primary at time $\tau \neq 0$ if and only if $\xi(\tau)=0$ and
$\phi(\tau) \in \{0,\pi\}$.  Then we must have $\tau=k\frac{ T_1}{2}=j
\frac{T_2}{2}$, with $k,j \in \Z\setminus \{0\}$.  Then
$\frac{k}{j}=\frac{m}{n}=q$ and this implies that there is $i \in \Z$
such that $k=im$ and $j=in$, because $(m,n)=1$.  It follows that
$\gamma(\tau)$ is a primary if and only if
$\tau=im\frac{T_1}{2}=in\frac{T_2}{2}=i\frac{T}{2}$, where
$T=mT_1=nT_2$ is the period of the orbit. This concludes the proof.
\qed
\end{proof}

If we assume $q=1$, then, using the above Proposition, we can exclude
at once the collision with the primaries for some simple cases.  To
simplify notation we place $\hat{A}_1(\beta)=\hat{A}_1(\beta,1)$.
\begin{proposition}
  \label{first_exceptions}
  Let $\gamma$ be a periodic orbit through the third centre $C$,
  corresponding to the value $\hat{A}_1(\beta)$, with $\beta<\beta_0$,
  as in Proposition~\ref{collisionC}. Let $(\xi_0,\phi_0)$ be fixed
  elliptic coordinates for the centre $C$. If $\phi_0=k\frac{\pi}{2},
  k \in \Z$, or $\xi_0=0$, then the orbit $\gamma$ cannot pass through
  the primaries.
\end{proposition}
\begin{remark}
  Note that the positions with $\phi_0=k\frac{\pi}{2}, k \in \Z$, and
  the ones with $\xi_0=0$ correspond in Cartesian coordinates $(x,y)$
  to points on the coordinate axes. In particular, if $\xi_0=0$ then
  the three centres are collinear on the $x$-axis, and $C$ lies
  between the primaries.  If $\phi_0=k\pi$ then the three centres are
  still aligned on the $x$-axis, but $C$ is external. Finally, if
  $\phi_0=(2k+1)\frac{\pi}{2}$, then the centre $C$ lies on the
  $y$-axis and the configuration of the centres is symmetric with
  respect to this axis.
  
  In particular, Proposition~\ref{first_exceptions} says that, for $q
  = 1$, the periodic orbits through the primaries intersect the
  coordinate axes only at the primaries and for $\xi=\xi_{\pm}$.
\end{remark}
\begin{proof} 
We are in the case $q=1$, then from Proposition~\ref{coll_primaries}
we know that the orbit $\gamma$ passes through a primary if and only
if for any time $\tau$ such that $\xi(\tau)=0$, we have
$\phi(\tau)=k\pi, k \in \Z$, and viceversa.  The centre $C$ does not
coincide with a primary, then it cannot happen that $\phi_0=k\pi, k\in
\Z$, and $\xi_0=0$ at the same time and in these cases the statement
is obvious.  Now suppose $\phi_0=(2k+1)\frac{\pi}{2}$.  The time
intervals from $C$ to a position with $\phi=i\pi, i \in \Z$, are
$(2j+1)\frac{T}{4}$, $j \in \Z$, where $T=T_1=T_2$ is the period of
$\gamma$. Look at the variable $\xi$: the only positions which have a
time distance of $(2j+1)\frac{T}{4}$ from $\xi=0$ are the inversion
points $\xi_{\pm}$, but $\xi_0\neq\xi_{\pm}$.
\qed
\end{proof}

\begin{remark} 
Note that the proof of Proposition~\ref{first_exceptions} cannot be
generalised to arbitrary fixed values of $q \in \Q^+$. For example, if
$q=1/2$, then, if a periodic orbit passes through $C_1$, it certainly
passes through a point with elliptic coordinates $(\xi,\pi/2)$, with
$\xi \in (0, \xi_+)$: in fact the time passed from the last passage
through $C_1$ to this point is $T_2/4=T_1/8$. Nevertheless, it's easy
to see that even for this case the passage through the primaries is
excluded when $\xi_0=0$ or $\phi_0=k\pi$: indeed, the time between two
positions with $\xi=0$ is $T_1/2=T_2$, and the points with
$\phi_0=k\pi$ along an orbit which passes through a primary must have
$\xi_0=0,\xi_{\pm}$.
\end{remark}

Our next step is to study the regularity of the function $\hat{A}_1(\beta,q)$ 
with respect to the real parameter $\beta$, to be used for proving the central 
result of the subsection: in particular, we are interested in the behaviour 
near $\beta=0$.
\begin{lemma}
  \label{regularA}
  Let $q \in \Q^+$ be fixed. 
  Then $\hat{A}_1$ is a smooth function of $\beta \in (0,1)$ and it can be 
  smoothly extended to $\beta=0$. In particular, there exists the limit 
  \[
  \hat{A}_1(0,q):=\lim_{\beta \rightarrow 0}\hat{A}_1(\beta,q) \in 
  \left(0,1\right)\,,
  \]
  the periods $T_1$ and $T_2$ are smooth for $\beta=0$ and 
  \[
  qT_1(0,\hat{A}_1(0,q))=T_2(0,\hat{A}_1(0,q))\ .
  \]
\end{lemma}
\begin{proof}
Denote by $F$ the function
\[
F(\beta, A_1):= [qT_1-T_2](\beta, A_1)\,,
\]
defined on the domain $\mathcal{D}=\left\{(\beta,A_1) \in \R^2| \quad
\beta \in (0,1)\,, \ 0<A_1<\frac{1}{1+\beta} \right\}$.

We observe that $\kappa_1, \kappa_2$ are $C^{\infty}$ functions of
$(\beta, A_1) \in \mathcal{D}$, and $\kappa_1,\kappa_2 \in (0,1)$;
then, by derivation under the integral sign, we conclude that $F$ is
$C^{\infty}$ on the same domain.  Furthermore $\frac{\partial
  F}{\partial A_1}\gneq 0$, then from the implicit function theorem we
can assert the regularity of $\hat{A}_1$ on $(0,1)$.

We want to extend the definition of the function $\hat{A}_1$ to
$\beta=0$.  First of all we prove that the function $F$ can be
smoothly extended to $\beta=0$.  We study $T_1$ and $T_2$ separately.
We start from $T_1$. If $\beta \in [0,1)$ and $A_1<\frac{1}{1+\beta}$,
  then $(1-4\beta A_1^2) \gneq 0$. It follows that $k_1^2$ is well
  defined and $C^{\infty}$ for $\beta =0$ and that $k_1^2 \in
  (\frac{1}{2},1)$. Then $T_1$ is $C^{\infty}$ for $\beta =0$.  Now
  see $T_2$. For $\beta \in [0,1)$, we have $\kappa_2^2 \in
    [0,\frac{1}{2})$ and it is a $C^{\infty}$ function of $\beta$.  To
      assert the regularity of $T_2$, it was enough to show
      $\kappa_2^2 < 1$, then we surely have $T_2$ defined and smooth
      for $\beta =0$.
        
The next step is to verify that for $\beta =0$ there is a unique value
$\hat{A}_1(0,q) \in (0, 1)$, such that $F(0,\hat{A}_1(0,q))=0$ and that
for this value $\frac{\partial F}{\partial A_1}(0,\hat{A}_1(0,q))\gneq
0$.  From the definition, we easily see that $\lim_{\kappa \rightarrow
  0}\K(\kappa)=\frac{\pi}{2}$, while $\lim_{\kappa \rightarrow
  1^-}\K(\kappa)=+\infty$. Moreover
\[
qT_1(0,A_1)= \frac{4q}{\sqrt{2a}} \K(\sqrt{\frac{A_1+1}{2}}) \,, \quad
T_2(0,A_1)= \frac{\pi}{\sqrt{a A_1}}\ .
\] 
$T_1(0,A_1)$ is a strictly increasing function of $A_1$, while
$T_2(0,A_1)$ is strictly decreasing, and
\[
\begin{array}{c}
\lim_{A_1\rightarrow 0} F(0,A_1)=-\infty \,, 
\\ 
\lim_{A_1 \rightarrow  1^-}F(0,A_1)=
\frac{1}{\sqrt{2a}} \left[\lim_{A_1 \rightarrow 1^-}
4q\K(\sqrt{\frac{A_1+1}{2}})- \sqrt{2}\pi\right] 
= +\infty \ .
\end{array}
\]
We conclude that $\hat{A}_1(0,q)$ is uniquely determined from the
equality $F(0,\hat{A}_1(0,q))=0$, and $\hat{A}_1(0,q) \in
(0,1)$. Moreover $\frac{\partial F}{\partial A_1}(0,\hat{A}_1(0,q))
\gneq 0$, then the regularity of the function $\hat{A}_1$ in $\beta
=0$ follows from the implicit function theorem.  \qed
\end{proof}

Now we are ready to state and show the main result.
\begin{theorem}
\label{dense_C}
Let $q \in \Q^+$ a fixed positive rational number.  There is a dense
open subset $X_q'\subset \R \times S^1$, such that fixed
$(\xi_0,\phi_0) \in X_q'$, there is $\beta_0>0$, such that for each
$\beta \in (0,\beta_0)$, the periodic orbits through $(\xi_0,\phi_0)$,
associated with $\hat{A}_1(\beta,q)$, do not pass through the
primaries $C_1,C_2$. In particular, after scaling time, they are
orbits of the not-regularised 2-centre problem with Lagrangian $L_0$,
and they have energy $E=-2a \beta \hat{A}_1$.
\end{theorem}
\begin{proof}
Let $(\xi_0,\phi_0) \in \R\times S^1$ be the position of the centre
$C$ in elliptic coordinates and let
$\gamma(\tau)=(\xi(\tau),\phi(\tau))$ be a periodic orbit associated
with $\hat{A}_1(\beta,q)$, which pass through $(\xi_0,\phi_0)$ with
velocity $(\xi'_0,\phi'_0)$. Without loss of generality we can assume
$\phi'_0>0$ (see Subsection~\ref{coll_C}).  Suppose that the orbit
$\gamma$ passes through a primary $C_1$ or $C_2$: at that instant we
have $\xi=0$ and $\phi \in \{0,\pi\}$.  Let $q=\frac{m}{n}$, with
$m,n\in \Z$ positive integers and $(m,n)=1$.  Let $\Delta\tau$ be the
shortest time interval to go from a primary to the centre $C$ along the
orbit $\gamma$. Thanks to Proposition~\ref{coll_primaries}, we must
have $\Delta\tau < \frac{T}{2}$, where $T=mT_1=nT_2$ is the period of
$\gamma$.

We have two possibilities, corresponding to start from $C_1$ or from
$C_2$.  The orbit solves the separated system~(\ref{sys_A}). Then, in
the first case
\[
\Delta \tau = \frac{1}{2\sqrt{a}}\int_0^{\phi_0}
\frac{1}{\sqrt{\beta\hat{A}_1\cos^2\phi+\hat{A}_1}} d\phi + iT_2 \,,
\]
while in the second case
\[
\Delta \tau = \frac{1}{2\sqrt{a}}\int_0^{\phi_0}
\frac{1}{\sqrt{\beta\hat{A}_1\cos^2\phi+\hat{A}_1}} d\phi 
\pm \frac{T_2}{2}
+ i'T_2 
\,,
\]
with $i,i' \in \Z$, $i,i'\geq 0$, and in the second case with sign $\pm$
respectively when  $\phi_0<\pi$, $\phi_0\geq\pi$. 

The condition $\Delta \tau < \frac{T}{2}=\frac{nT_2}{2}$ implies that 
$i<\frac{n}{2}$ and $i'<\frac{n+1}{2}$. 
Look now at the variable $\xi$. In both cases we must have
\[
\Delta \tau = \pm \frac{1}{2\sqrt{a}} \int_0^{\xi_0}
\frac{1}{\sqrt{\cosh \xi - \beta \hat{A}_1 \cosh^2 \xi - \hat{A}_1}}d\xi 
+j\frac{T_1}{2}
\ .
\]
with $j \in \Z$, $j\geq 0$,
and the condition $\Delta \tau < \frac{T}{2}=\frac{mT_1}{2}$ implies 
that $j \leq m$.

We denote by $P(\phi_0,\beta)$ and $Q(\xi_0,\beta)$ the functions
\begin{eqnarray*}
P(\phi_0,\beta)&=& \frac{1}{2\sqrt{a}}\int_0^{\phi_0}
\frac{1}{\sqrt{\beta\hat{A}_1\cos^2\phi+\hat{A}_1}} d\phi 
\,,
\\
Q(\xi_0,\beta)&=& \frac{1}{2\sqrt{a}} \int_0^{\xi_0}
\frac{1}{\sqrt{\cosh \xi - \beta \hat{A}_1 \cosh^2 \xi - \hat{A}_1}}d\xi 
\ .
\end{eqnarray*}
From Lemma~\ref{regularA} and the fact that $\xi_0$ is not an inversion point, 
we deduce that these functions are smooth in a neighbourhood of $\beta=0$.
The condition to pass through a primary for the first case is
\[
\Delta \tau
= P(\phi_0,\beta) + i T_2 
= \pm Q(\xi_0,\beta) + j \frac{T_1}{2}
\,,
\]
while for the second case it is 
\[
\Delta \tau
= P(\phi_0,\beta) \pm \frac{T_2}{2}+ i' T_2 
= \pm Q(\xi_0,\beta) + j \frac{T_1}{2}
\ .
\]
We have a finite set of possible values for the integers $i, i',j$ and the 
equality $qT_1=\frac{m}{n}T_1=T_2$ holds, then we can summarise all the 
conditions with the following one:
if the orbit $\gamma$ passes through a primary, then 
$G^+(\xi_0,\phi_0,\beta) \in S$ or $G^-(\xi_0,\phi_0,\beta) \in S$, where 
$S \subset \Q$ is a finite subset of rationals, depending only on the fixed 
parameter $q \in \Q^+$, and the functions $G^+,G^-$ are defined by
\[
G^{\pm}(\xi_0,\phi_0,\beta)= 
\frac{P(\phi_0,\beta)\pm Q(\xi_0,\beta)}{T_1(\beta, \hat{A}_1(\beta,q))}
\ .
\]
The functions $G^{\pm}$ are smooth in all the variables: in
particular, if $G^{\pm}(\xi_0,\phi_0,0)\notin S $, there is
$\beta_0>0$, such that $G^{\pm}(\xi_0,\phi_0,\beta) \notin S$ for any
$\beta \in (0,\beta_0)$.  Then, to exclude the collision with the
primaries for small values of $\beta$, it is enough that the centre
$C$ belongs to the set
\[
X_q'=\{(\xi_0,\phi_0)\in \R \times S^1| 
\quad G^{\pm}(\xi_0,\phi_0,0)\notin S \}\ .
\] 
The set $S$ is finite, then it is clear that $X_q'$ is open and dense
in $\R \times S^1$. Indeed, we easily see from the definitions of the
functions $G^{\pm}$ that their partial derivatives with respect to
$\phi_0$ are different from zero; then $X_q'$ is a finite intersection
of open dense subset of $\R \times S^1$.  Moreover, the complement of
$X_q'$ has zero Lebesgue measure.  \qed
\end{proof}

\begin{remark}
When $q=1$, it follows easily from Proposition~\ref{coll_primaries}
that at least one of the two periodic orbits through $C$ corresponding
to the same $\hat{A}_1(\beta)$ does not collide with the primaries.
Indeed, without loss of generality we can assume $\phi'_0>0$ (see the
end of Subsection~\ref{coll_C}).  Suppose $\phi_0 \in
(0,\frac{\pi}{2})$: then the time to pass from the centre $C_1$ to the
centre $C$ must be less than $\frac{T}{4}$, where $T=T_1=T_2$ is the
period. If $\xi_0>0$ the only possibility to collide with the
primaries is that $\xi'_0>0$, while if $\xi_0<0$ we must have
$\xi'_0<0$.  For the other possible intervals of values of $\phi_0$ a
similar reasoning works.

For our purposes, the fact of having at least one orbit through $C$
that does not collide with the primaries is not enough: indeed, after
we have obtained the collision arcs, we want to construct
\emph{collision chains} with them, and to do it we need at least two
orbits of the system $(L_0)$ with the same energy $E$, which pass
through $C$ with not-parallel tangent fields.  The latter condition
will be investigated in the next section.
\end{remark}

\begin{remark}
Theorem~\ref{dense_C} would be improved if we showed one of the
following:
\begin{itemize}
\item[i)] the partial derivatives $\frac{\partial G^{\pm}}{\partial
  \beta}(\xi_0,\phi_0,0)$ are zero only for isolated values of
  $(\xi_0,\phi_0)$;
\item[ii)] the partial derivatives $\frac{\partial G^{\pm}}{\partial
  \beta}(\xi_0,\phi_0,0)$ do not vanish for every $\xi_0 \in \R$,
  $\phi_0 \in S^1$, with $(\xi_0,\phi_0)$ not a primary.
\end{itemize}
For case (i) we would have that the collision with the primaries is
possible only for isolated positions $(\xi_0,\phi_0)\in \R \times
S^1$, while for case (ii) the collision would be excluded for any
position of the centre $C$.

At present, we don't have any of these improvements.
\end{remark}

\begin{corollary}
\label{dense_C_entire}
Let $I \subset \Q^+$ be a finite set of positive rationals.  There is
a dense open subset $X_I' \subset \R \times S^1$, such that for each
$(\xi_0,\phi_0)\in X_I'$, there is $\beta_0>0$, such that for any
$\beta \in (0,\beta_0)$, the periodic orbits through $(\xi_0,\phi_0)$
corresponding to $\hat{A}_1(\beta,q)$, with $q \in I$, do not pass
through the primaries. In particular, they are periodic orbits for the
system $(L_0)$.
\end{corollary}
\begin{proof}
It is an immediate consequence of the construction of the subset $X_q'
\subset \R \times S^1$ in the proof of Theorem~\ref{dense_C}: we can apply
this theorem for any $q \in I$, then take the intersection of the sets
$X'_q$ thus obtained, and the resulting set maintains the same
properties of the sets $X'_q$.  \qed
\end{proof}

\begin{remark}
\label{XX_I}
Denote by $\psi: \R \times S^1 \rightarrow \R^2$ the map to pass from
elliptic to Cartesian coordinates, $\psi(\xi,\phi) = \cosh(\xi+i\phi)
= x+iy$.  The map $\psi$ is open and surjective then, fixed a
finite subset $I \subset \Q^+$, the image $X_I=\psi(X_I')$ is an open
dense subset of $\R^2$.  Then we can say that the thesis of
Corollary~\ref{dense_C_entire} holds for any position of the centre
$C$ in an open dense subset $X_I \subset M$.
 
Consider now the set $X'=\cap_{q \in \Q^+}X_q'$: by the classical
Baire's Category Theorem the set $X'$ is dense in $\R \times S^1$.
Then, the set $X=\psi(X')$ is dense in $\R^2$ and for any finite
subset $I\subset \Q^+$, it is contained in $X_I$.  Then we can say
that the thesis of Corollary~\ref{dense_C_entire} holds for any
position $C \in X \subset M$.  In this manner the only information
that we have lost is that we don't know if the set $X$ is open, while
we have gained the independence of the dense set $X$ of positions of
$C$ from the set of rationals $I$.  Note that the choice of a
sufficiently small $\beta$ cannot be independent of $I$, instead.
\end{remark}

\subsection{Early collisions}
\label{early_coll}
We defined a collision arc in Section~\ref{bolotin} to be a critical
point of the Maupertuis-Jacobi functional which starts and ends at
collision with the centre $C$ and does not meet this centre at
intermediate times.  If we take one of the periodic orbits through $C$
found in the preceding section, and we pass to Cartesian coordinates,
then we cannot be sure that the orbit does not pass newly through $C$
before a period has passed.  This can happen in two ways: when the
orbit, considered in elliptic coordinates, meets the point
$(\xi_0,\phi_0)$ and when it meets $(-\xi_0,-\phi_0)$.

When a periodic orbit starting from the centre $C$ passes newly
through $C$ in a time shorter than its period, then we talk of
\emph{early collision}.  To understand when an early collision occurs
we need to study the behaviour of periodic orbits a little deeper.

An easy example is given by the case $q=1$: in this case any periodic
orbit crosses the $y$-axis, is symmetric with respect to this axis
and, if it does not collide with the primaries, it has a transverse
autointersection at the point of crossing.  The symmetry comes from
the fact that in a time interval of half the orbit's period we have
the passage from a point $(\xi,\phi)$ to a point $(-\xi,\phi +\pi)$,
that is from a point $(x,y)=(\cosh (\xi) \cos (\phi), \sinh (\xi) \sin
(\phi))$, to its symmetric $(-x,y)$.  The intersections with the
$y$-axis occur when $\phi=\pm \pi/2$ and from one intersection to the
next there is a time interval of half the period.  In elliptic
coordinates the orbit passes through a point $(\xi, \pi/2)$ and after
half a period it arrives at $(-\xi,-\pi/2)$, but these two points
coincide in Cartesian coordinates and are on the $y$-axis. Then each
orbit autointersects at a point on the $y$-axis.  The transformation
of the velocities when we pass from elliptic to Cartesian coordinates
is given by:
\begin{equation}
\label{matrixU}
v=U(\xi,\phi) \cdot \left(
\begin{array}{c}
\xi' \\ \phi'
\end{array} \right)\,, \quad
U(\xi,\phi) =
\left(
\begin{array}{cc}
\sinh \xi \cos \phi & \ -\cosh \xi \sin \phi
\\
\cosh \xi \sin \phi &  \ \sinh \xi \cos \phi
\end{array}
\right)\,,
\end{equation}
where $U(\xi,\phi)$ is an invertible matrix, except when $(\xi,\phi)
\in \{(0,0),(0,\pi)\}$.  We observe that $U(-\xi,-\phi)=-U(\xi,\phi)$.
If at the point of intersection with the $y$-axis $\xi = \pm \xi_+$,
then $\xi'=0$ and the velocity has zero $y$-component: then the two
crossings are not transverse.  But this is the case in which the orbit
collides with the primaries, reversing its direction at the
collisions.  If $\xi \neq \pm \xi_+$ instead, the velocities at the
point of intersection with the $y$-axis have both the components
different from zero, then they are transverse, because by symmetry
they differ only in the sign of their $x$-component.

\begin{figure}[h]
\begin{minipage}[b]{0.49\textwidth} 
  \centering
   \includegraphics[width=\textwidth]{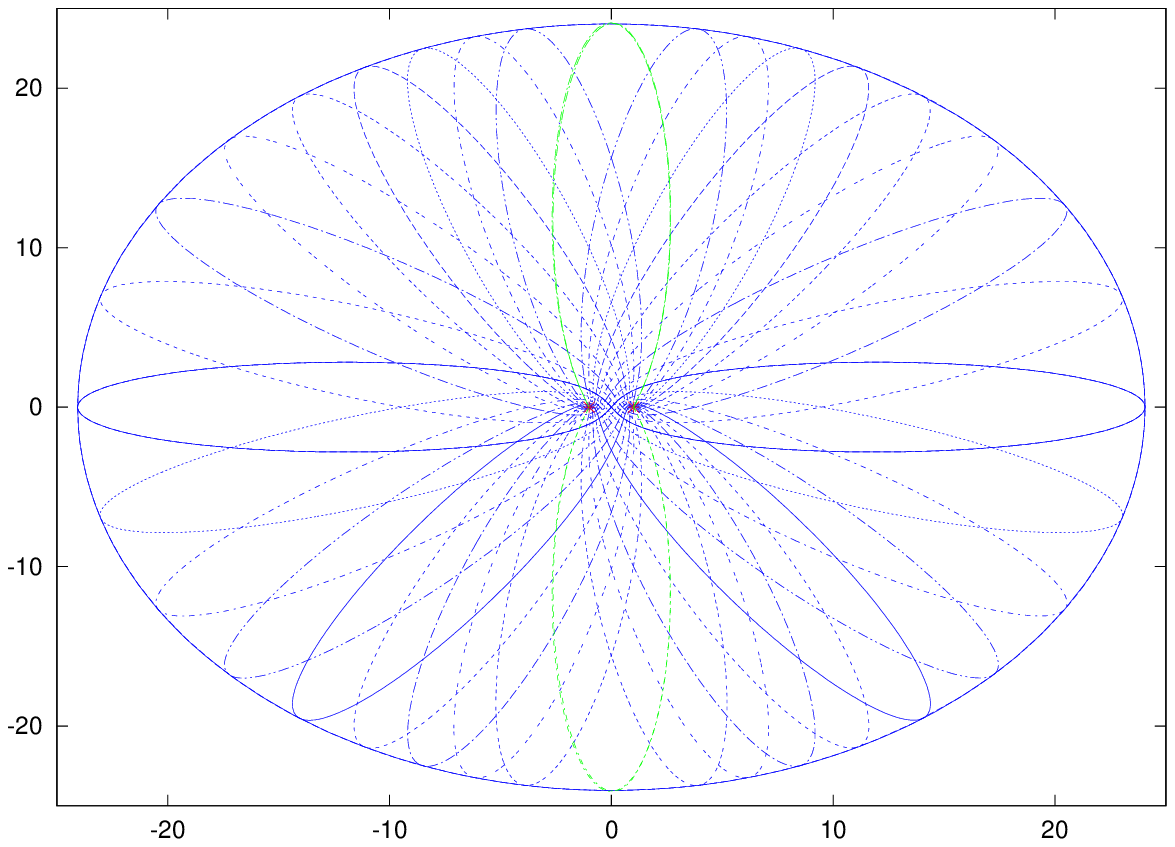}
   \caption{\small{Periodic orbits in Cartesian coordinates when $a=1$, $q=1$, 
     $\beta=1/7$ and the orbit through the primaries. }}
   \label{perorb1}
 \end{minipage}
 \hspace{2mm} 
 \begin{minipage}[b]{0.49\textwidth}
  \centering
  \includegraphics[width=\textwidth]{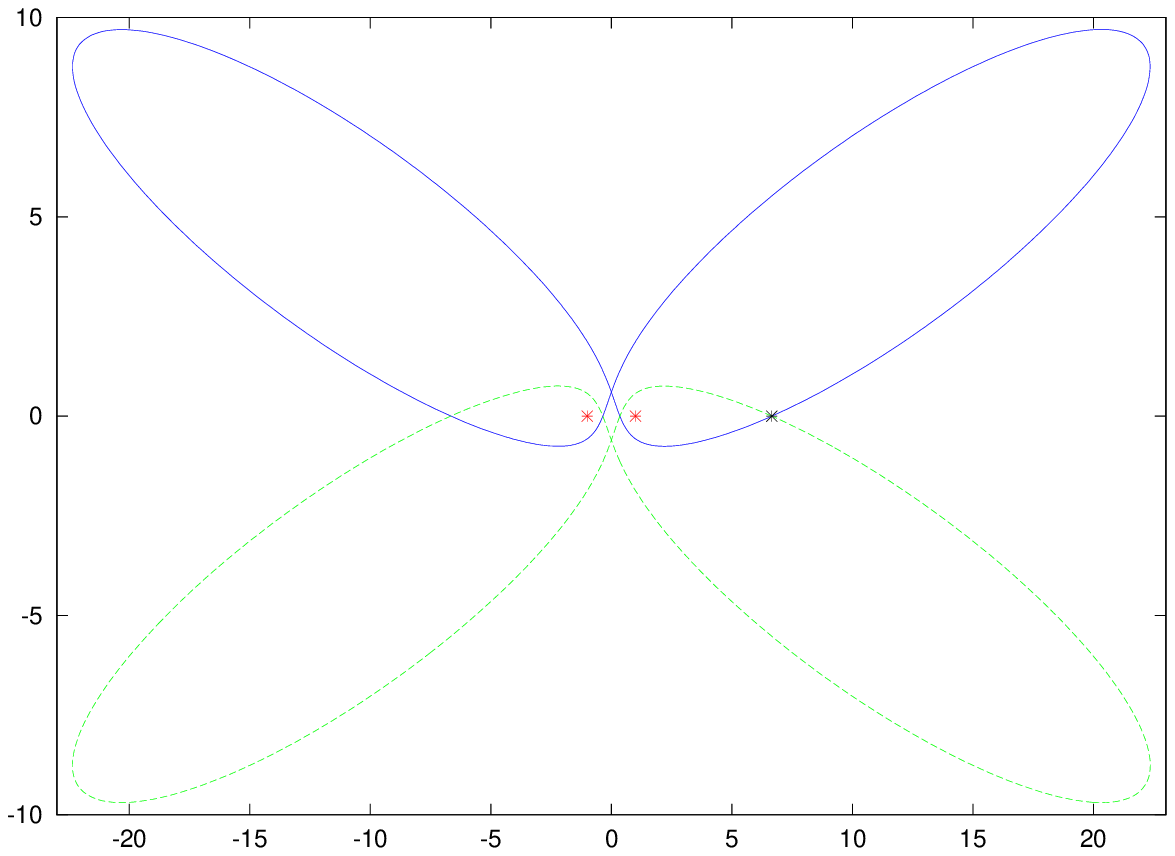}
   \caption{\small{Periodic orbits through
       $(\xi_0,\phi_0)=(\frac{2}{3} \xi_+,0)$ in Cartesian
       coordinates, when $a=1$, $q=1$, $\beta=1/7$.}}
   \label{perorb2}
 \end{minipage}
\end{figure}

An illustration of this situation is given in Figure~\ref{perorb1},
where we have drawn $18$ periodic orbits corresponding to the value of
$\beta=1/7$, included the two orbits that collide with the primaries:
these latter orbits are in green, while the primaries are marked with
red asterisks.  We have also drawn the periodic orbit $\xi=\xi_+$: we
cannot use it to construct collision arcs, because it has different
energy, but this orbit is interesting in itself because it encloses
all the trajectories with the same values of the parameters (and then
the same energy), and it is tangent to all.  The behaviour of a couple
of periodic orbits through the same point can be seen in
Figure~\ref{perorb2}, where the starting point is marked with a black
asterisk.

Then, if $q=1$, we can say that there is an early collision at $C$ if and only
if $C$ lies on the $y$-axis: in this case there is a unique, up to reversing 
the direction of motion, periodic orbit through $C$.

In particular, we see that the period does not change in the passage
from elliptic to Cartesian coordinates.  Actually, this holds for any
value of the parameter $q \in \Q^+$.  In fact, if we take a periodic
orbit that starts from a point $(\xi_0,\phi_0)$, which is not a
primary, we see that the matrix $U(\xi_0,\phi_0)$ given in
(\ref{matrixU}) is invertible and
$U(-\xi_0,-\phi_0)=-U(\xi_0,\phi_0)$.  Then the only possibility to
have a shorter period when we pass to Cartesian coordinates is that
the orbit passes through $(-\xi_0,-\phi_0)$, with velocity
$(-\xi_0',-\phi_0')$.  This is clearly impossible because the sign of
$\phi'$ never changes along the orbit.
 
\begin{figure}[h]
 \begin{minipage}[b]{0.49\textwidth}
   \centering
   \includegraphics[width=\textwidth]{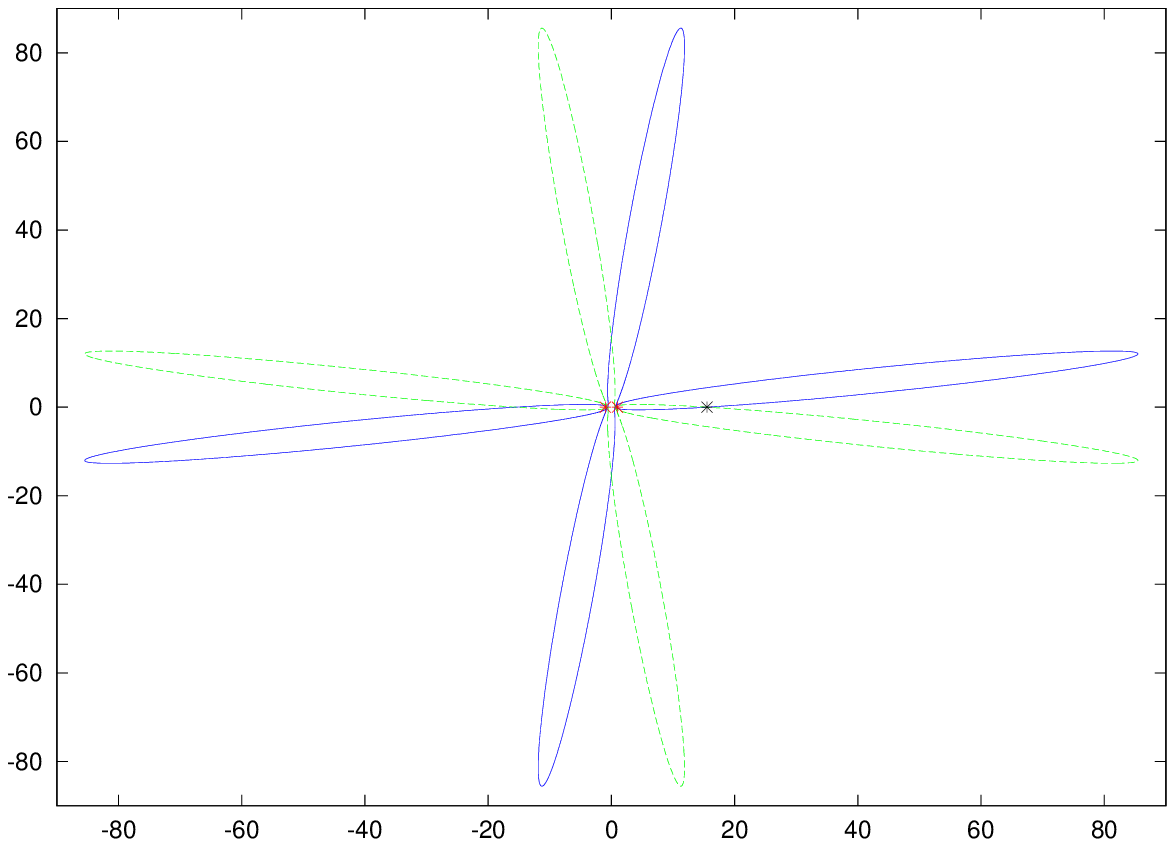}
   \caption{\small{Periodic orbits through 
       $(\xi_0,\phi_0)=(\frac{2}{3} \xi_+,0)$ in Cartesian coordinates, 
       when $a=1$, $q=2$, $\beta=1/7$. }}
   \label{perorb3_3}
 \end{minipage}
 \hspace{2mm} 
 \begin{minipage}[b]{0.49\textwidth}
  \centering
  \includegraphics[width=\textwidth]{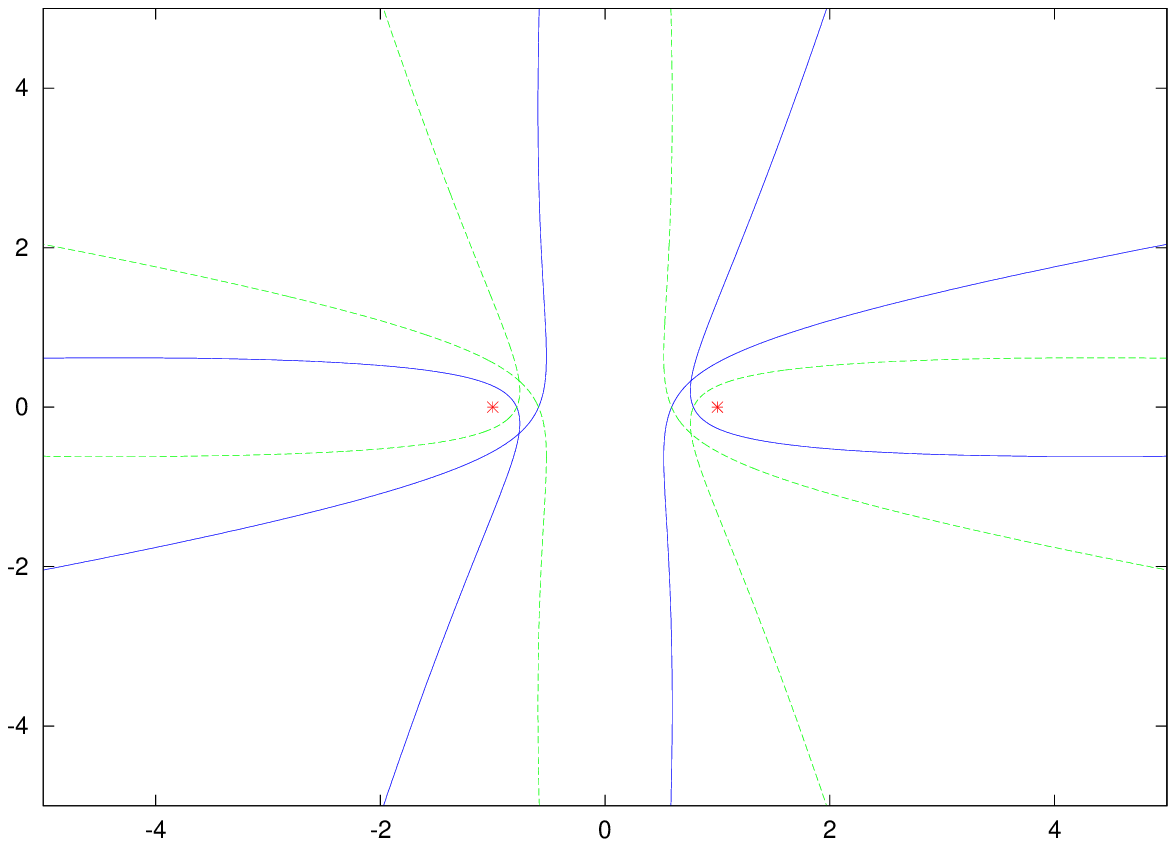}
   \caption{\small{An enlargement of Figure~\ref{perorb3_3}, where the 
       autointersections of the two orbits are visible.}}
   \vspace{0.06cm}
   \label{perorb3_3ingrand}
 \end{minipage}
\end{figure}

Anyway, the situation complicates if $q \neq 1$ and we cannot get a global 
view easily: as an example, in  Figure~\ref{perorb3_3} we have drawn the 
orbits through the point $(\xi_0,\phi_0)=(\frac{2}{3} \xi_+,0)$, when $a=1$,
$\beta=1/7$ and $q = 2$.
To better see the autointersections we have put an enlargement in  
Figure~\ref{perorb3_3ingrand}.

We note the following remarkable fact:
\begin{proposition}
\label{ellip_early_coll}
Let $(\xi_0,\phi_0) \in \R \times S^1$ be fixed and consider a
periodic orbit $(\xi(\tau),\phi(\tau))$ through this point,
corresponding to some fixed values of $\beta,q$.  If
$(\xi(\tau),\phi(\tau))$ passes newly through $(\xi_0,\phi_0)$ before
closing, then it is the unique periodic orbit through $(\xi_0,\phi_0)$
corresponding to the same fixed values of $\beta, q$, up to reverse
the direction of motion. Moreover, if $\xi_0$ is not an inversion
point, the orbit has a transverse auto\newtheorem{Theorem}{Theorem}intersection at
$(\xi_0,\phi_0)$.
\end{proposition}
\begin{proof}
If the initial velocity in $(\xi_0,\phi_0)$ is $(\xi_0',\phi_0')$,
then the only other possible velocity through the same point along the
same orbit is $(-\xi_0',\phi_0')$.  If the orbit arrives at
$(\xi_0,\phi_0)$ before that a period is passed, then it must happen
with velocity $(-\xi_0',\phi_0')$, and this implies that the two
possible periodic orbits through $(\xi_0,\phi_0)$ coincide.  \qed
\end{proof}

At this stage, we know completely early collisions only for the case
$q=1$.  In general, we can only say that early collisions cannot be
excluded: we don't have any knowledge about the conditions that
determine them.  Anyway, it is not a problem for the proof of
Theorem~\ref{central_result}.  When an early collision occurs, we will
take as collision arc the partial arc of the periodic orbit through
$C$, which starts from $C$ and ends at the first next passage through
$C$.  The final time $T$ of the collision arc in this case will not be
the period, but it will be the time of the first return to $C$.

Finally, we want to stress that a collision arc must end at $C$ with one 
of the four possible velocities determined by the choice of the parameters,
which in elliptic coordinates are 
$\{\pm (\xi_0',\phi_0'),\pm (-\xi_0',\phi_0')\}$,
where $(\xi_0',\phi_0')$ is the initial velocity of the arc.

\section{Nondegeneracy and direction change.}
\label{degen_dir_change}
In this section we will verify that the collision arcs obtained from 
Corollary~\ref{dense_C_entire} satisfy the nondegeneracy condition and 
moreover they meet transversely at the centre $C$: this will allow us to 
apply Theorem~\ref{bol-mack} and derive our final result, 
Theorem~\ref{central_result}.  

\subsection{Nondegeneracy}
\label{nondegen}
Let $q \in \Q^+$ be fixed and $m,n \in \Z$ be positive coprime
integers such that $q=m/n$.  Denote by $(\xi_0,\phi_0) \in \R\times
S^1$ the position of the centre $C$ in elliptic coordinates.  Suppose
that $(\xi_0,\phi_0) \in X_q'$, where the set $X_q' \subset \R \times
S^1$ is the one given by Theorem~\ref{dense_C}. Take $\beta$ small
enough, so that the periodic orbits of system $(\mathcal{L}_0)$
corresponding to $\hat{A}_1(\beta,q)$ and passing through $C$ do not
pass through the primaries.  Let $\gamma(t)$ be one of the resulting
collision arcs for the system $(L_0)$, which starts from the centre
$C$ with velocity $v_0$ and whose energy is $E=-2a \beta
\hat{A}_1(\beta,q)$. Then $\gamma(t)$ solves the system
\begin{equation}
\label{sys_degen}
f(C,v_0,T)=C\,, \quad H_0(C,v_0)=E\ .
\end{equation}
 As reminded in Section~\ref{bolotin}, to show the nondegeneracy of
 $\gamma$ it is sufficient to verify that the Jacobian of
 system~(\ref{sys_degen}) is nonzero. Actually we will verify a slight
 variant of this condition.

It is convenient to consider as variables the parameters $(\beta,
A_1)$ instead of the coordinates of the initial velocity $v_0$. This
procedure is right only if there is a local diffeomorphism which
allows to pass from $v_0$ to $(\beta,A_1)$. The transformation of the
velocities in the passage from elliptic to Cartesian coordinates is
given by the invertible matrix $U(\xi_0,\phi_0)$ defined in
(\ref{matrixU}).  We know that the orbit $\gamma$ corresponds to a
solution of the separated system~(\ref{sys_A}), and in elliptic
coordinates we have
$(\dot{\xi}_0,\dot{\phi}_0)=\frac{d\tau}{dt}(0)(\xi'_0,\phi'_0)$, with
$\frac{d\tau}{dt}(0) = (\cosh^2(\xi_0) - \cos^2(\phi_0))^{-1}$. Then
$v_0$ is a $C^{\infty}$ function of $(\xi'_0,\phi'_0)$. From the
system~(\ref{sys_A}) we see that $(\xi'_0,\phi'_0)$ is a $C^{\infty}$
function of $(\beta,A_1)$ and then $v_0$ is. By the same reasoning we
see that $(\beta,A_1)$ is locally a $C^{\infty}$ function of the
velocity $(\xi'_0,\phi'_0)$ and then of $v_0$.

For all nearby trajectories from the same initial point $C$, we
evaluate the $m$-th positive instant of time at which they meet the
ellipse $\xi=\xi_0$, with the velocity $\dot{\xi}$ equal to the
initial velocity $\dot{\xi}_0 \neq 0$; then we consider the time
distance from the $n$-th passage through $\phi_0$, which is given by
$nF(\beta,A_1)$, where
\[
F(\beta,A_1)=[qT_1-T_2](\beta,A_1)\ .
\] 
This is equivalent to consider the value of the angle coordinate
$\phi$ at the instant at which we have the $m$-th \emph{oriented}
crossing of the line $\xi=\xi_0$.

Note that in this manner the time variable $T$ is fixed as function of
$(\beta, A_1)$, $T=mT_1(\beta,A_1)$, so that we have reduced the order
of system~(\ref{sys_degen}) of a unit.  The orbit $\gamma$ satisfies
\[
F(\beta,A_1)=0\,, \quad -2a \beta A_1=E\,,
\]
and it is nondegenerate if the Jacobian determinant of this system at
the solution $(\beta,\hat{A}_1(\beta,q))$, corresponding to the orbit
$\gamma$, is different from zero.

The Jacobian matrix is 
\[ J=
\left(
\begin{array}{cc}
\frac{\partial F}{\partial \beta} & 
\frac{\partial F}{\partial A_1} \\
-2a A_1 &
-2a \beta
\end{array}
\right)\ .
\]
As observed in Subsection~\ref{per_orb}, we have $\partial F/\partial
A_1 \gneq 0$. Furthermore, we have that $\hat{A}_1(\beta,q) \in
(0,\frac{1}{1+\beta})$. These properties remain true for $\beta =0$,
as proved in Lemma~\ref{regularA}, and moreover $T_1,T_2$ are
$C^{\infty}$ functions of $(\beta,A_1)$.  It follows that when $\beta
=0$ and $A_1=\hat{A}_1(0,q)$, the determinant is well defined and
different from zero, and by regularity the same is true also for
values $\beta,\hat{A}_1(\beta,q)$, with $\beta>0$ sufficiently small.

We conclude that the collision arc $\gamma$ is nondegenerate for $\beta$  
small enough.

\subsection{Direction change}
\label{dir_change}
Let $q \in \Q^+$ be fixed and $m,n \in \Z$ be positive coprime
integers such that $q=m/n$.  In order to construct \emph{collision
  chains} to which apply Theorem~\ref{bol-mack}, we should have at
least two collision arcs which start and arrive at the centre $C$ with
transverse tangent fields.

From the results of Section~\ref{collision}, if in elliptic
coordinates our centre $(\xi_0,\phi_0)$ belongs to the dense open
subset $X_q' \subset \R \times S^1$, then for small enough values of
$\beta$, after reparametrisation of time and change of coordinates,
the orbits associated to $\hat{A}_1(\beta,q)$ do not collide with the
primaries and they are periodic orbits for the not-regularised
two-centre problem $(L_0)$.  As observed in Subsection~\ref{coll_C},
there are exactly two orbits corresponding to $\hat{A}_1(\beta,q)$,
which pass through the centre $(\xi_0,\phi_0)$ with velocities
$(\xi'_0,\phi'_0)$, $(-\xi'_0,\phi'_0)$ respectively.  Changing sign
to $\phi'_0$, we obtain simply the same orbits, with reversed
direction of motion.  Moreover, the two transverse orbits coincide in
the case of an autointersection at $(\xi_0,\phi_0)$, as we have seen
in Proposition~\ref{ellip_early_coll}.

Now consider the two orbits associated with $\hat{A}_1(\beta,q)$:
suppose they pass through the centre $(\xi_0,\phi_0)$ at time $\tau=0$
and that their velocities are $(\xi'_0,\phi'_0)$,
$(-\xi'_0,\phi'_0)$. They are obviously transverse at the point
$(\xi_0,\phi_0)$ in the cylinder $\R \times S^1$, because we have
chosen $\beta$ small enough to have $\xi'_0\neq 0$ (see
Proposition~\ref{collisionC}).  We must verify that this
transversality is conserved after time reparametrisation and changing
from elliptic to Cartesian coordinates.  The reparametrisation of time
is given by formula (\ref{newtime}): it maintains the directions,
because the centre $(\xi_0,\phi_0)$ is not a primary.  As seen in
Subsection~\ref{early_coll}, the passage to Cartesian coordinates is
given by an invertible matrix $U$ (see (\ref{matrixU})), then the
transversality is conserved.

Look now at the other possible elliptic coordinates for $C$: they are
$(-\xi_0,-\phi_0)$, then the possible velocities at this point are the
same as the ones at $(\xi_0,\phi_0)$. It follows that there are no
more orbits through $C$ that we can consider, for the same values of
$\beta, q$.  We conclude that there are two transverse directions for
the collision arcs starting from $C$. In correspondence of the same
values of the parameters $\beta, q$ we obtain four collision arcs,
divided in pairs of arcs with transverse initial velocities.

Finally, we observe that, in correspondence of some fixed values of
$\beta, q$, if an early collision occurs at $C$, then there is a
unique (up to reverse the direction of motion) periodic orbit through
$C$ with transverse autointersection at $C$. Indeed, since the passage
through the primaries is excluded, the matrix $U$ is invertible and we
cannot have inversion points along the orbit, because $\phi' \neq 0$.

We summarise the results of this and the preceding subsection in the
following
\begin{proposition}
\label{transversality}
Let $I \subset \Q^+$ be a finite set of positive rationals.  Suppose
that the centre $C$ has elliptic coordinates $(\xi_0,\phi_0) \in
X_I'$, where $X_I' \subset \R\times S^1$ is the dense subset given by
Corollary~\ref{dense_C_entire}. Then, there is $\beta_0>0$ such that
for each $\beta \in (0,\beta_0)$ and for each $q \in I$, there are
exactly four collision arcs at $C$ for the system $(L_0)$, associated
with the value $\hat{A}_1(\beta,q)$, which are nondegenerate.
Moreover, the four collision arcs divide into two pairs, according to
their initial velocities at $C$: any couple is formed by two arcs with
opposite initial velocities at $C$, and each arc in one pair has
transverse initial velocity to each arc in the other.  \qed
\end{proposition}

\subsection{Proof of Theorem~\ref{central_result}}
\label{final_proof}
We're going to conclude the proof of Theorem~\ref{central_result}.

Let $I \subset \Q^+$ be a finite set of positive rationals and  
$X_I=\psi(X_I')\subset M$ as in Remark~\ref{XX_I}.
The energy of the collision arcs at $C$ is given as function of the parameters 
$(\beta,q)$, $q \in I$, by the relation
\[
E=-2a \beta \hat{A}_1(\beta,q)\ .
\]
Thanks to the regularity of the function $\hat{A}_1(\beta,q)$ for 
$\beta \in [0,1)$ and to the fact that $\hat{A}_1(0,q)\in (0,1)$ 
(Lemma~\ref{regularA}), to require $\beta$ small enough is equivalent to ask 
for the absolute value of the energy $E$ to be small enough. In particular, 
if $\beta<\beta_0$, with $\beta_0$ sufficiently small, then the energy 
$E(\beta,q)$ is a strictly decreasing function of $\beta$: 
\[
E(\cdot,q):(0,\beta_0) \rightarrow (-E_0,0)\,,
\] 
with $E(0,q)=0$, $E(\beta_0,q)=-E_0<0$.  It follows that all the
previous results can be stated using the energy parameter $E$ instead
of $\beta$.

If we choose the energy $E<0$ sufficiently close to zero, 
then for each $q \in I$, there are four  nondegenerate collision arcs 
of energy $E$ through the centre $C$: they are pieces of 
periodic trajectories on the configuration space 
$M=\R^2 \setminus\{C_1,C_2\}$.

For $\beta$ fixed, the energy increase with the class $q$ (see 
Proposition~\ref{A1_respect_q}). This means that we 
cannot state a general result valid for a fixed energy $E$ and any $q\in \Q^+$. 
On the other hand, we don't need such a result, because to apply 
Theorem~\ref{bol-mack}   we only want a finite number of collision arcs. 
What is certainly true is that for any finite set of classes $I\subset \Q^+$ 
of cardinality $i$, we can choose a small enough energy value $E<0$ to form 
a set of $4i$ collision arcs of energy $E$, by taking for each $q\in I$ the 
four arcs obtained by our procedure. 

By the monotony of the function $\hat{A}_1(q,\beta)$ with respect to
$q$ and its relation with the energy $E$, we are sure that the arcs
with the same energy $E$, but different classes, cannot have the same
value of $\hat{A}_1$. Then the velocities $\xi_0',\phi_0'$ cannot
coincide (see system~(\ref{sys_A})).  However, we can't state that
collision arcs of different classes determine a different set of
directions at the point $C$. What we can certainly assure is that if
we fix an arc of class $q\in I$, then for each class $q' \in I$, not
necessarily different from $q$, we can always choose a pair of arcs of
class $q'$, which start at $C$ with directions transverse to the
velocity with which the arc of class $q$ has arrived at $C$.
    
Thus, for any fixed sequence $\{q_k\}_{k \in \Z}$, $q_k \in I$, we can
construct infinite collision chains $\{\gamma_k\}_{k \in \Z}$, such
that any $\gamma_k$ is a piece of a periodic orbit of the $2$-centre
problem $(L_0)$ of class $q_k$.
 
The assumptions of Theorem~\ref{bol-mack} are all satisfied and we can apply 
this theorem and then Theorem~\ref{bol-mack_hyperb}
to get our main result Theorem~\ref{central_result}, 
which is now definitely proved.
\qed

\vspace{1cm}
{\bf Acknowledgements }
I wish to thank P.~Negrini for having introduced me to the subject.

\appendix
\begin{footnotesize}
\section{Positivity of entropy}
\label{app:entropy}

In this appendix we give a proof of the positivity of the topological
entropy of the Poincar\'e map for the case at hand.  We do it by
computing the exponential growth rate of the number of periodic
orbits.

By local uniqueness, if $(\gamma_{k_i})_{i \in \Z}$ is a periodic
collision chain, then the shadowing orbit is also periodic. We
compute for each positive integer $n$ the number $P_n$ of periodic
collision chains with period $n$ and take the limit
\[
p=\limsup_{n \rightarrow \infty} \frac{\log (P_n)}{n}\ .
\]

Let $q$ be fixed and consider only the collision arcs of class $q$.
They are exactly four, corresponding to the two possible transverse
directions at the centre $C$. We distinguish two cases:
\begin{itemize}
\item[i)] there are exactly two transverse periodic orbits of the
  unperturbed problem through $C$ (up to reverse the direction of
  motion), and the arcs are given by the entire periodic orbits;
\item[ii)] there is only one periodic orbit of the unperturbed problem
  through $C$ (up to reverse the direction of motion), with a
  transverse autointersection at $C$, and the arcs correspond to parts
  of the periodic orbit between two successive passages through $C$.
\end{itemize}

In case i), for a sequence $(\gamma_{k_i})_{i \in \Z}$ to be a
collision chain we must have that $\dot \gamma_{k_i}(0)=\dot
\gamma_{k_i}(T_{k_i})\neq \pm \dot \gamma_{k_{i+1}}(0)$. Then we
cannot have periodic orbits with odd period $n=2m+1$, $m \in \Z^+$ and
$P_{2m+1}=0$ for each $m \in \Z^+$. The number of periodic orbits with
even period is $P_{2m}=2^{2m+1}$ instead, then $p=\log 2$.

In case ii), for $(\gamma_{k_i})_{i \in \Z}$ to be a collision chain,
we must have that $\dot \gamma_{k_i}(0)=\pm \dot \gamma_{k_{i+1}}(0)$,
because there are only four possible velocities at $C$, divided in two
pairs each containing the parallel ones, and any arc $\gamma_{k_i}$
arrives at $C$ with velocity $\dot \gamma_{k_i}(T_{k_i})$ transverse
to the initial one $\dot \gamma_{k_{i}}(0)$. Then $P_n=2^{n+1}$, for
any $n \in \Z^+$, and again $p=\log 2$.

We conclude that the entropy corresponding to any choice of $I \subset
\Q^+$ is $h_{\rm top} \geq \log 2$.


\end{footnotesize}

\end{document}